\documentclass[prb,reprint]{revtex4-2}
\usepackage{graphicx,amssymb,amsfonts,amsmath}
\usepackage{color}
\usepackage{mathtools}
\usepackage{latexsym}
\usepackage{graphicx}
\usepackage{float}
\usepackage{dcolumn}
\usepackage{bm}
\usepackage{amssymb}
\usepackage{amsmath}
\usepackage{mathtools}
\usepackage[space]{grffile}
\usepackage{xcolor}
\usepackage{comment}

\usepackage{hyperref}
\hypersetup{
    colorlinks=true,
    linkcolor=blue,
    filecolor=blue,      
    urlcolor=blue,
    citecolor=blue
    }

\begin{document}

\title{Decay rates of almost strong modes in Floquet spin chains beyond Fermi's Golden Rule}
\author{Hsiu-Chung Yeh$^{1}$}
\author{Achim Rosch$^{2}$}
\author{Aditi Mitra$^{1}$}
\affiliation{
$^{1}$Center for Quantum Phenomena, Department of Physics,
New York University, 726 Broadway, New York, NY, 10003, USA\\
$^{2}$Institute for Theoretical Physics, University of Cologne, 50937 Cologne, Germany
}

\begin{abstract}
The stability and dynamics of almost strong zero and $\pi$ modes in weakly non-integrable Floquet spin chains are investigated. Such modes can also be viewed as localized Majorana modes at the edge of a topological superconductor. Perturbation theory in the strength of integrability-breaking interaction $J_z$ is employed to estimate the decay rates of these modes, and compared to decay rates obtained from exact diagonalization. 
The structure of the perturbation theory and thus the lifetime of the modes is governed by the conservation of quasi-energy modulo $2 \pi/T$, where $T$ is the period of the Floquet system. If there are $4 n-1$ bulk excitations whose quasi-energies add up
to zero (or $\pi/T$ for a $\pi$ mode), one obtains a decay channel of the Majorana mode with a decay rate proportional to $J_z^{2 n}$.
Thus the lifetime is sensitively controlled by the width of the single-particle Floquet bands. For regimes where the decay rates are quadratic in $J_z$, an analytic expression for the decay rate in terms of an 
infinite temperature autocorrelation function of the integrable model is derived, and shown to agree well with exact diagonalization. 
\end{abstract}
\maketitle
\section{Introduction}

The study of topological phenomena is a frontier topic in condensed matter physics, with the classification of topological phases showing further enrichment under periodic or Floquet driving \cite{OkaRev,SondhiRev}. At the single-particle level, topological phenomena are well understood, with the defining characteristics being a bulk boundary correspondence, where a non-zero bulk topological invariant predicts the number of edge modes.  However, the effect of weak interactions on the stability of edge modes in topological systems in general, and Floquet topological systems in particular, is not fully understood. With Floquet driving and interactions, the system is susceptible to heating, and one might naively expect all edge physics to quickly melt away on the same time-scales as typical bulk observables heat. Nevertheless, if the system hosts strong modes  \cite{Kitaev01,Fendley2012,FendleyXYZ,Alicea16,Fendley17,Nayak17,Garrahan18,Garrahan19,Yates19,Yates20a,Yao20,Yates20,Yates21,Yates22,Bardarson22,Fendley23}, typically encountered at the edge of certain one-dimensional systems, the edge modes can be stable over time scales much longer than bulk heating times \cite{Fendley17,Nayak17,Yates19,Yates20,Yao20,Yates20a,Yates21,Yates22}.

For Floquet systems with $Z_2$ symmetry, strong modes are of two kinds, zero modes that are local Majorana operators that commute with a Floquet unitary, and $\pi$ modes that are local Majorana operators that anti-commute with the Floquet unitary \cite{Zoller11,Sen13,Else16,Roy16,Khemani16, Potter16,Yates19}. Since this property is not tied to any eigenstate of the system, any Floquet eigenstate with open boundary conditions, can show signatures of these edge modes. These strong modes also anti-commute with the $Z_2$ symmetry leading to eigenspectrum phases. In particular, the zero mode implies each eigenstate is at least doubly degenerate with the degenerate pairs being even and odd under parity ($Z_2$ symmetry), while the $\pi$ mode implies every even parity eigenstate has an odd parity partner whose quasi-energy is larger than the former by $\pi/T$ (modulo $2\pi/T$), where $T$ is the Floquet period.

So far, all Floquet spin chains for which strong modes have been derived can be represented as unitaries of fermion bilinears \cite{Sen13,Yates19}. Besides an exact construction of the operators, strong modes can also be detected by studying the infinite temperature autocorrelation function of an operator on the edge \cite{Fendley17,Yates19,Yates20,Yao20,Yates20a}. If a strong mode exists, and the operator has an overlap with it, then the infinite temperature autocorrelation function is long lived, with a lifetime that grows exponentially with system size \cite{Fendley2012,FendleyXYZ}. This approach can easily be extended to non-integrable Floquet spin chains \cite{Yates22,Yates21}, where the non-integrability is via the application of unitaries of four-fermion interactions. In this case, one expects that the correlation function does ultimately decay but possibly after a very long time. 

Employing the diagnostic of the autocorrelation function, obtaining the interaction dependent lifetime can be notoriously difficult, especially when the integrability breaking is very weak  \cite{yeh2023Impurity}. This is because, for weak integrability breaking, the decay is primarily due to finite size effects, with the system sizes needed to be in a regime where the decay is controlled purely by the integrablity breaking terms, being much too large. For this reason, for weakly non-integrable systems, these edge  modes are referred to as almost strong modes (ASM) \cite{Fendley17,Nayak17,Yates19,Yates20a,Yao20,Yates20}. In this paper we explore to what extent perturbation theory  can help estimate the decay times of almost strong modes of weakly non-integrable Floquet spin chains. We identity regimes where Fermi's Golden Rule (FGR) is valid, and also regimes where FGR breaks down and higher powers of the integrability breaking term are needed to capture the decay.

The paper is organized as follows. We introduce the model in Section \ref{sec:Model}, summarizing the essential properties in the integrable limit, and introducing the infinite temperature autocorrelation function. In Section \ref{sec:FGR} we derive the FGR decay rates in terms of an appropriate infinite temperature autocorrelation function of the integrable model. We also highlight regimes where FGR breaks down and we present a simple argument to determine what powers of the integrability breaking term control the decay. In Section \ref{sec:Results} we present the numerical results and compare them to theoretical predictions. We present our conclusions in Section \ref{sec:Conc}, while intermediate steps in the derivations are relegated to the appendices. 

\section{Model} \label{sec:Model}
We study stroboscopic time-evolution of an open chain of length $L$ according to the Floquet unitary 
\begin{align}
    U = e^{-i\frac{T}{2}J_z H_{zz}}e^{-i\frac{T}{2}g H_z}e^{-i\frac{T}{2}J_x H_{xx}},
    \label{Eq: Full Unitary}
\end{align}
where 
\begin{align}
    H_{xx} = \sum_{i=1}^{L-1} \sigma_i^x\sigma_{i+1}^x;\ H_{z} = \sum_{i=1}^{L} \sigma_i^z;\ H_{zz} = \sum_{i=1}^{L-1} \sigma_i^z\sigma_{i+1}^z.
    \label{Eq: Ising Hamiltonian}
\end{align}
Above $\sigma^{x,y,z}_i$ are Pauli matrices on site $i$,  $g$ is the strength of transverse-field and $J_{x,z}$ is the strength of the Ising interaction in the $x,z$-direction. We will set $J_x = 1$ in the following discussion. $T$ qualitatively represents the period of the drive. In particular, taking $T\ll 1$ would recover the high frequency limit where the Floquet Hamiltonian is simply $H_F =i(\ln{U})/T = J_z H_{zz}/2 + g H_z/2 + H_{xx}/2$.

For $J_z$=$0$, the Floquet unitary becomes non-interacting, $U_0 = U|_{J_z = 0}$, and the corresponding  Floquet Hamiltonian $H_F^0 = i(\ln U_0)/T $, can be solved analytically for periodic boundary conditions. In particular (see Appendix \ref{sec:A}) we find
\begin{align}
    H_F^0 = \sum_{k>0} \epsilon_k \left( d_k^\dagger d_k - d_{-k}d_{-k}^\dagger \right),
    \label{Eq: Free Floquet Hamiltonian}
\end{align}
where $d_k^\dagger (d_k)$ are the creation (annihilation) operators of Bogoliubov quasi-particles, and the bulk dispersion $\epsilon_k$ is
\begin{align}
    \cos(\epsilon_k T) =  \cos(gT)\cos(J_xT) + \sin(gT)\sin(J_x T)\cos{k}.
    \label{Eq: Bulk Dispersion}
\end{align}
The quasi-energy band $\epsilon_k$ becomes exactly flat when either $J_x T=0, \pi$ or $g T= 0, \pi$. The consequence of this on  the decay-rate of the edge modes, will be emphasized later.

In contrast to continuous time, the quasi-energy is only defined modulo $2\pi/T$, $\epsilon_k T \in [-\pi,\pi]$.  Since within the Bogoliubov formalism, for each state with energy $\epsilon$ there has to be a state with energy $-\epsilon$, two values of $\epsilon$, $\epsilon=0$ and $\epsilon=\pi/T$ are special. This is because they have the property that $\epsilon T=-\epsilon T \mod 2 \pi$, allowing for topologically protected modes at those energies. Therefore, two different types of edge modes, zero mode $\psi_0$ and $\pi$ mode $\psi_\pi$ can exist, with these modes appearing or disappearing via the gap closing at $0$ or $\pi$. Depending on the choice of parameters, $g$ and $T$, the system may possess none, one or both of the two edge modes. When these edge modes exist, they anti-commute with the $Z_2$ symmetry, $\mathcal{D} = \sigma_1^z \ldots \sigma_L^z$, of the system: $\{ \psi_0, \mathcal{D} \} = \{ \psi_\pi, \mathcal{D} \}$. In the thermodynamic limit of a semi-infinite chain, the zero and $\pi$ modes respectively commute and anti-commute with the Floquet unitary, $[\psi_0 ,U_0] = 0$, $\{ \psi_\pi, U_0 \} = 0$. Due to this property, they have infinite lifetime in the thermodynamic limit. Appendix \ref{sec:B} presents exact expressions for the $0$ and $\pi$ modes for the Floquet Ising model $U_0$. 

\begin{figure}[h!]
    \centering    \includegraphics[width=0.45\textwidth]{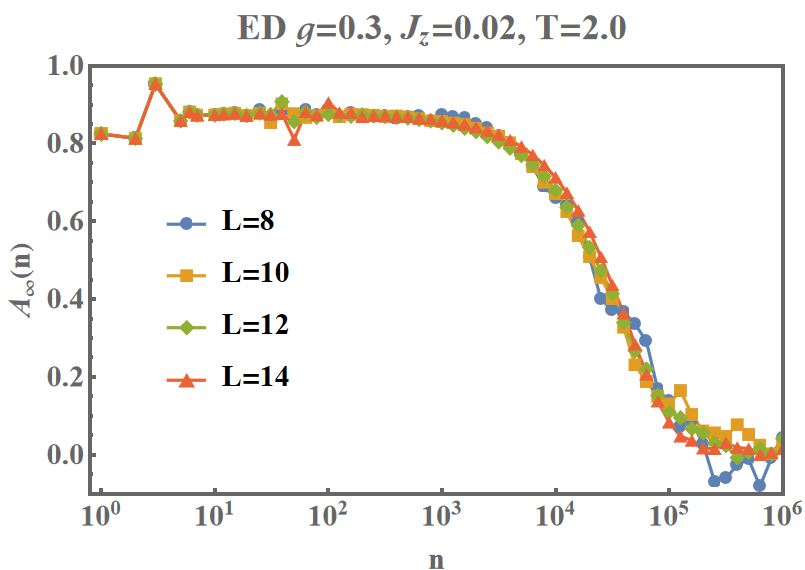} \includegraphics[width=0.45\textwidth]{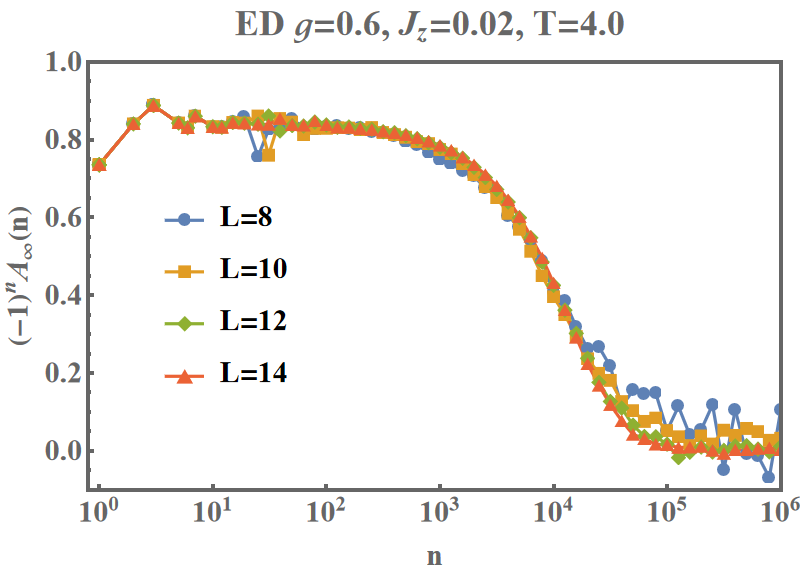}
    \caption{The infinite temperature correlation $A_\infty$ from exact diagonalization for different system sizes $L = 8, 10, 12, 14$. The plots show signatures of an almost strong zero mode (top panel) and an almost strong $\pi$ mode (bottom panel) that survives for many cycles $n$ before decaying. Note the different microscopic parameters $g,T$ for the panels.}
    \label{Fig: Correlation example}
\end{figure}

When interactions are turned on, the commutation (anti-commutation) relations between zero ($\pi$) mode and the Floquet unitary is violated. However, these edge modes are still long-lived quasi-stable modes, and are known as ASM. A useful quantity to probe this phenomena is the  autocorrelation of $\sigma_1^x$
\begin{align}
    A_\infty (n) = \frac{1}{2^L} \text{Tr}[\sigma_1^x(n) \sigma_1^x],
    \label{Eq: AutoCorrelation Sigmax}
\end{align}
where $n$ is the stroboscopic time-period. This is a good measure of the almost strong mode dynamics in the presence of interactions since both zero and $\pi$ modes are localized on the edge with $\mathcal{O}(1)$ overlap with $\sigma_1^x$, $\text{Tr}[\psi_0 \sigma_1^x]/2^L \sim \mathcal{O}(1)$ and $\text{Tr}[\psi_\pi \sigma_1^x]/2^L \sim \mathcal{O}(1)$. In the language of Majoranas, $\sigma_1^x$ is the Majorana on the first site and the edge modes are a superposition of Majoranas, with largest weight being on the Majoranas near the boundary. We show an example of $A_\infty(n)$ in Fig. \ref{Fig: Correlation example}. 

Fig.~\ref{Fig: Correlation example} shows examples of ASMs. The top panel is for
$g=0.3, T=2.0$ and shows an almost strong zero mode. The bottom panel is for
a longer period $T=4.0$ and a larger transverse-field $g=0.6$, and shows an almost strong $\pi$ mode. Both panels are for the
same strength of the integrability breaking term $J_z =0.02$.
After an initial transient, the autocorrelation decays into a long-lived zero $\psi_0$ (upper panel) or $\psi_\pi$ mode (lower panel) depending on the value of parameters $g,T$. The modes survive for many periods, indicated by the  constant value of the autocorrelation function for the zero mode and the extra oscillation with the period $(-1)^n$ for the $\pi$ mode. Eventually, the autocorrelations decay to zero due to interactions. Here the decay occurs around $\approx 10^4$ time cycles for both modes. Autocorrelation functions of bulk quantities (with no overlap with the ASMs) decay much faster, within $1$-$2$ drive cycles. 

For our model, the decay of ASMs can be captured by perturbation theory in the integrability breaking term. The interaction enables the scattering between the edge mode and bulk Majorana operators, with the leading contribution to the decay arising from resonance conditions, i.e., conditions that determine when the energy of the edge mode matches the total energy of certain number of bulk excitations, modulo $2\pi/T$. In the next section, we derive the decay rate of the infinite temperature autocorrelation function within second order perturbation theory i.e, $O(J_z^2)$, equivalently, FGR. Subsequently, we also discuss resonance conditions which allow us to predict the leading power of $J_z$ controlling the decay rates. 

\section{FGR  and beyond for infinite temperature autocorrelation function} \label{sec:FGR}

Let us first examine the almost strong zero mode. In general, one can decompose the Floquet unitary into two parts
\begin{align}
   U &= e^{-i V T} e^{-i H_F^0 T},
\end{align}
where $V$ is the perturbation applied to the system and $H_F^0$ is the Floquet Hamiltonian of the unperturbed system. For the case we study in ~\eqref{Eq: Full Unitary}, one can identify the perturbation with $V = J_z H_{zz}/2$ and $H_F^0 = i(\ln U_0)/T$, where $U_0 = \exp(-iTg H_z/2)\exp(-iTJ_x H_{xx}/2)$. Performing a second order expansion of $V$, the Floquet unitary is 
\begin{align}
    U \approx \left(1 - i V T - \frac{V^2 T^2}{2} \right)e^{-i H_F^0 T}.
\end{align}
After a period, the zero mode evolves into $\psi_0(1)= U^\dagger \psi_0 U$, and to second order in
$V$, it is given by 
\begin{align}
    \psi_0(1) \approx e^{i \mathcal{L}_0T}\left(1 + iT\mathcal{L}_V + T^2\mathcal{G}_{V^2} - \frac{T^2}{2}\mathcal{F}_{V^2}\right)\psi_0.
\end{align}
Above, we have used the notation: $\mathcal{L}_0 \psi_0 = [H_F^0, \psi_0],\ \mathcal{L}_V\psi_0  = [V,\psi_0],\ \mathcal{G}_{V^2}\psi_0  = V \psi_0 V$ and $\mathcal{F}_{V^2} \psi_0 = \{V^2,\psi_0 \}$. Notice that $e^{i \mathcal{L}_0T} \psi_0 = U_0^\dagger \psi_0 U_0 = \psi_0$ comes from the commutation relation between the zero mode and the unperturbed Floquet unitary. 

After $N$ periods we obtain (see Appendix \ref{sec:C})
\begin{align}
    \psi_0(N) = \left[e^{i\mathcal{L}_0T}\left(1 + iT\mathcal{L}_V + T^2\mathcal{G}_{V^2} - \frac{T^2}{2}\mathcal{F}_{V^2}\right)  \right]^N \psi_0.
    \label{Eg: zero mode N period}
\end{align}
Now, we are in the position to calculate the autocorrelation of the zero mode
\begin{align}
    A_\infty^0 (N) = \frac{1}{2^L} \text{Tr} [\psi_0(N) \psi_0 ].
\end{align}
By inserting \eqref{Eg: zero mode N period} into the autocorrelation $A_\infty^0$ and keeping terms up to second order in $V$, one arrives at
\begin{align}
    &A_\infty^0 (N) \nonumber\\ &= A_\infty^0 (0) -
    \frac{N T^2}{2^L} \biggl( \frac{1}{2}  \text{Tr} \left[ \dot{\psi_0} \dot{\psi_0} \right] + \sum_{n\geq 1}^{\infty } \text{Tr}\left[\dot{\psi_0}(n) \dot{\psi_0}\right] \biggr),
\end{align}
where we define $\dot{\psi_0} = i\mathcal{L}_V\psi_0$, $\dot{\psi_0}(n) = e^{i \mathcal{L}_0nT}\dot{\psi_0}$ and  $A_\infty^0 (0)=1$ in our case.

\begin{figure*}[t]
    \centering    \includegraphics[width=0.21\linewidth]{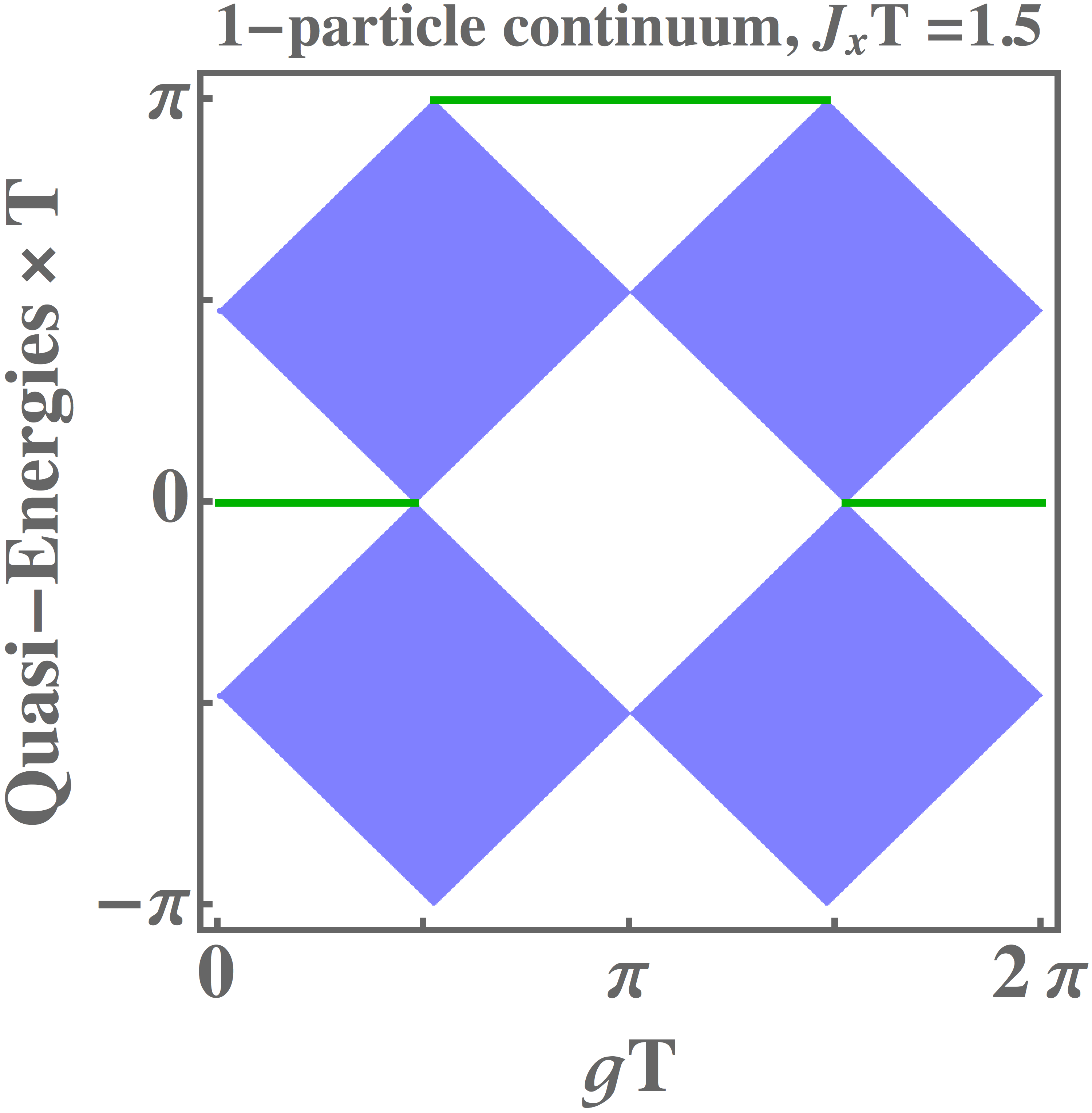}\hspace{0.1cm}
                 \includegraphics[width=0.21\linewidth]{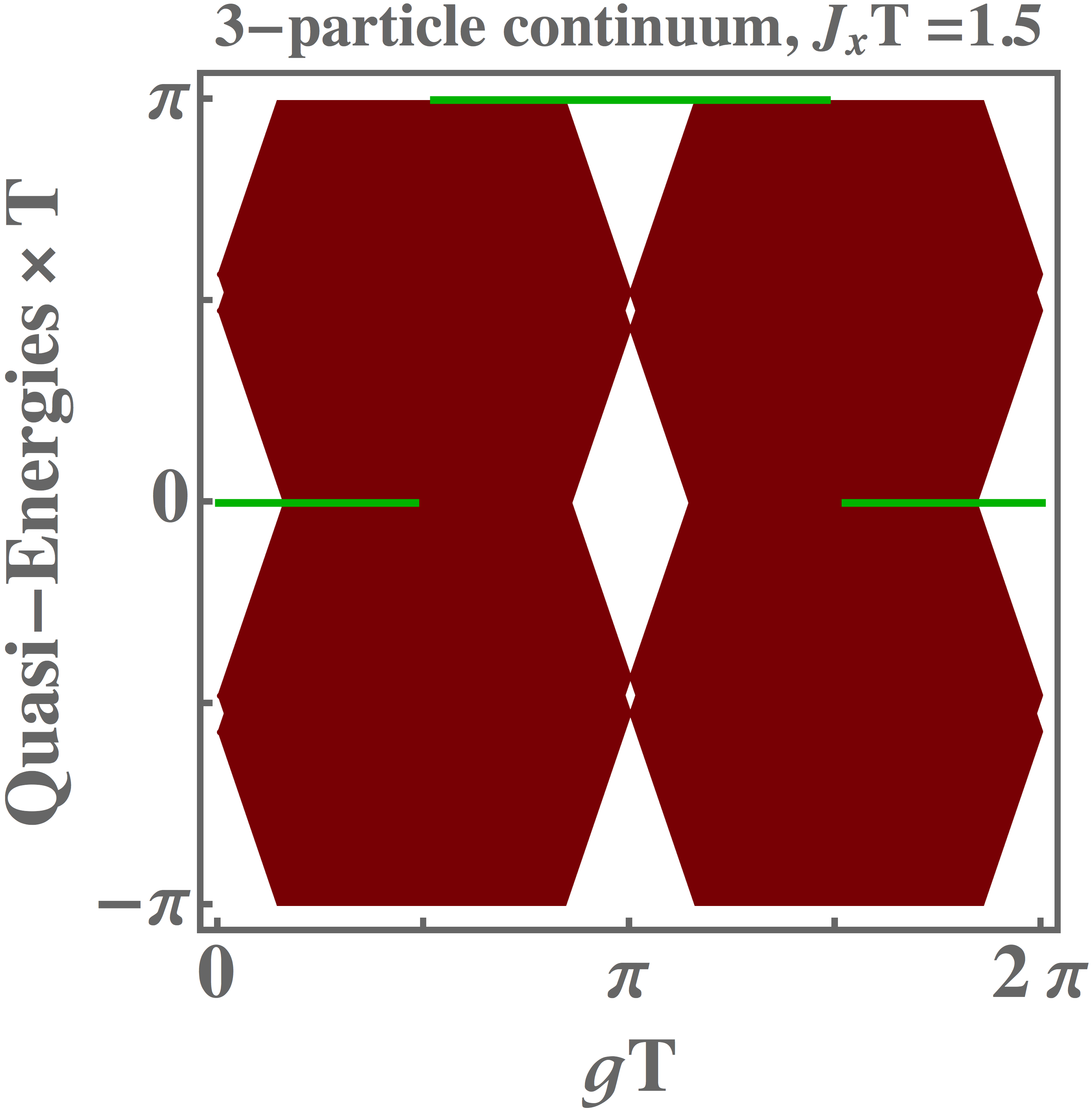}  \hspace{0.1cm}
                  \includegraphics[width=0.21\linewidth]{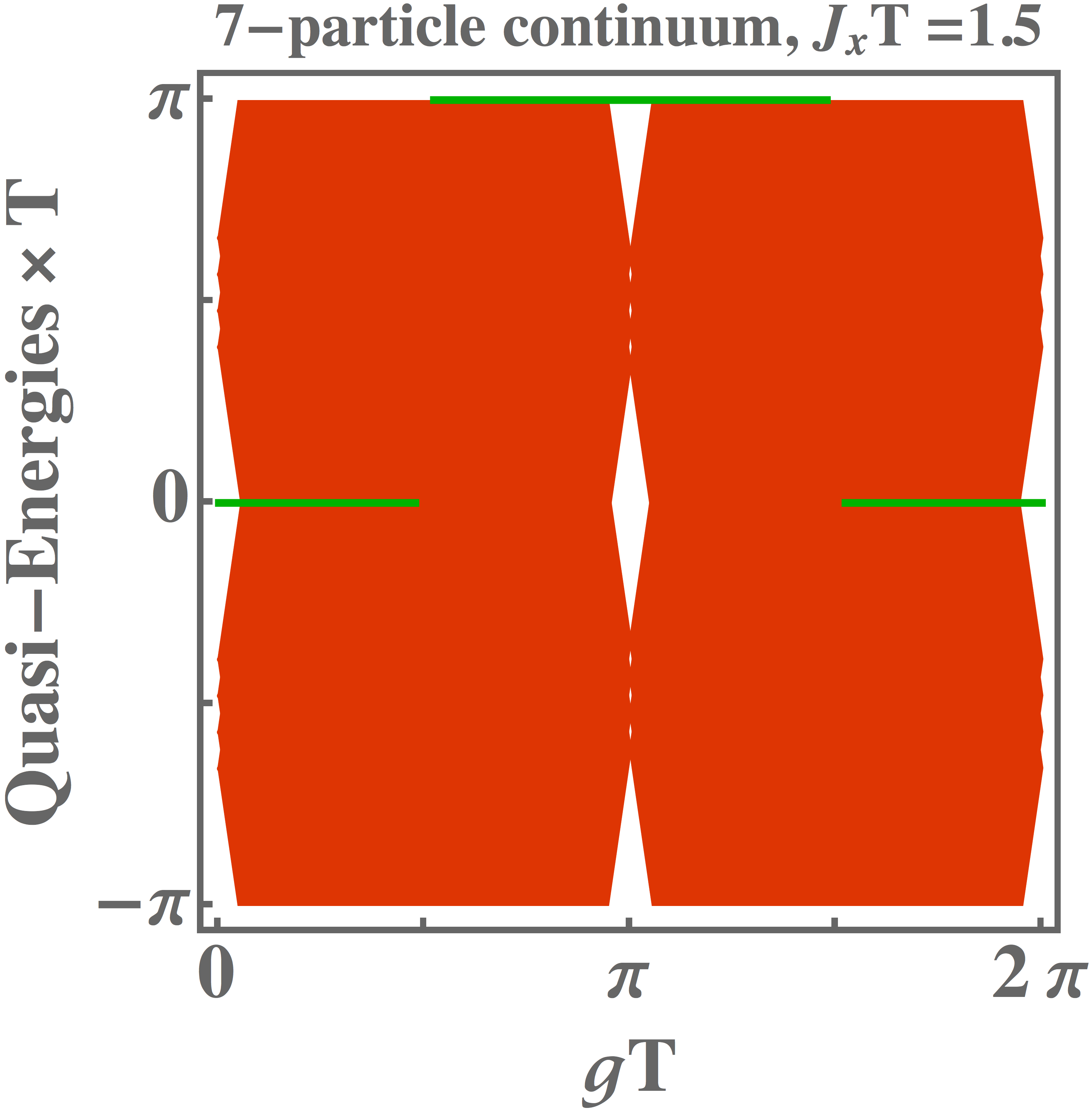}  \hspace{0.1cm} \includegraphics[height=0.215\linewidth]{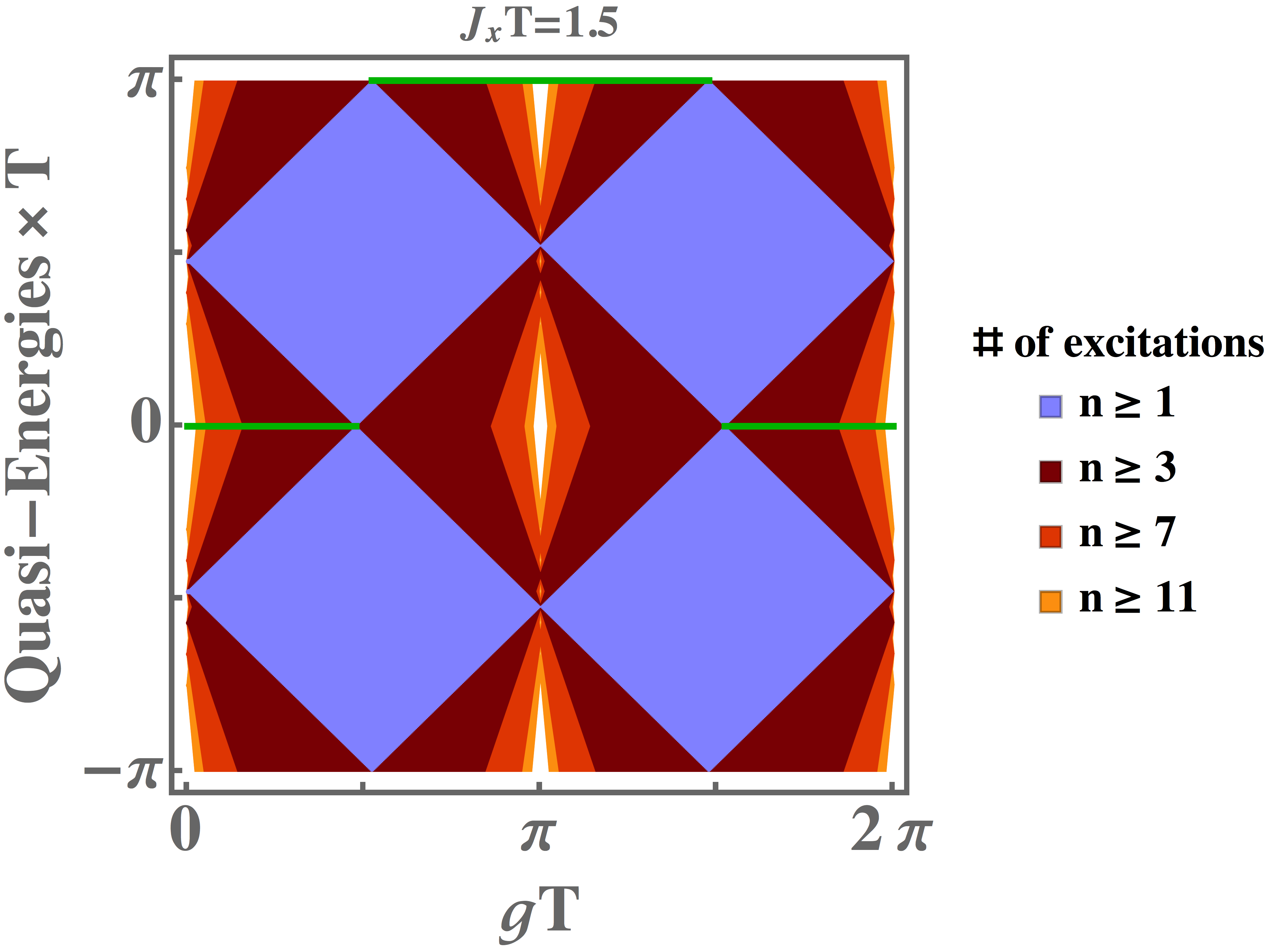} 
    \caption{Plot of quasi-energies of $n$-particle excitations, $n=1, 3, 7$, and a plot, where all these continua are combined, for $J_x T=1.5$, and as a function of $g T$. See Appendix \ref{sec:D} for a description of how the continua are constructed. The thick dark green lines denote the position of the Majorana modes. The plot can be used to identify the dominant scattering process leading to a decay of the almost strong modes, see text.}
    \label{Fig:continua}
\end{figure*}

The autocorrelation $A_\infty^0(N)$ with decay rate $\Gamma_0$ can be approximated by $A_\infty^0(N) \approx A_\infty^0(0)  \exp(-N T \Gamma_0) \approx A_\infty^0(0) (1 - N T \Gamma_0)$ within perturbation theory. From this we conclude (see details in Appendix \ref{sec:C}) that the FGR decay rate for the zero mode correlation at second order in the perturbation, and in the large $N$ limit is, 
\begin{align}
    \Gamma_0 &= \frac{T}{ A_\infty^0 (0) 2^L} 
    \biggl( \frac{1}{2}  \text{Tr} \left[ \dot{\psi_0} \dot{\psi_0} \right] + \sum_{n= 1}^{\infty} \text{Tr}\left[\dot{\psi_0}(n) \dot{\psi_0}\right] \biggr) \nonumber\\
    &= \frac{1}{A_\infty^0 (0)2^L} \sum_{i,j} |\langle i| \Tilde{V} |\Tilde{j}\rangle|^2  \pi \delta_F\left(\epsilon_i -\epsilon_j \right).
    \label{Eq: Fermi's Golden Rule 0Mode}
\end{align}
$| i \rangle$ are the many-particle eigenstates of the 
unperturbed unitary $U_0$ with eigenvalue $e^{-i \epsilon_i T}$. We define $\Tilde{V} = V - \psi_0 V \psi_0 $ and $|\Tilde{j} \rangle = \psi_0 |j\rangle$. Using $\psi_0^2=1$, one can rewrite $\Tilde{V} =[V,\psi_0] \psi_0$, thus $\tilde V$ can be interpreted as the part of the interaction which does not commute with the zero mode and thus changes it. Above, the $\delta_F$ function encodes energy conservation modulo $2 \pi/{T}$, with $\delta_F(\epsilon)=\sum_m \delta(\epsilon+m 2\pi/T)$. In contrast to the traditional FGR for time-independent Hamiltonians, here one obtains quasi-energy conservation, rather than energy conservation, due to the Floquet time evolution.

 The FGR formula \eqref{Eq: Fermi's Golden Rule 0Mode} has been derived from a short-time expansion, $N T \Gamma_0 \ll 1$, and it is therefore strictly speaking only valid if short and long-time behaviors are governed by the same decay process. This question has been studied in the context of the memory matrix formalism applied to integrable systems \cite{Jung2006,Jung2007} and integrability breaking perturbations \cite{Jung2007b}. The analysis confirms that the perturbative formula is valid provided that the investigated mode is the slowest mode in the system: in this case short- and long-time decay coincide. This is justified for the almost strong modes studied in this paper.

For the $\pi$ mode, the only difference comes from the anti-commutation relation, $e^{i \mathcal{L}_0T} \psi_\pi = U_0^\dagger \psi_\pi U_0 = -\psi_\pi$. It leads to extra factors of $(-1)$ in the derivation. One can show (see Appendix \ref{sec:C}) that the $\pi$ mode decay rate at second order in the perturbation is 
\begin{align}
    \Gamma_\pi &= \frac{T}{A_\infty^\pi (0)2^L} \left( \frac{1}{2}  \text{Tr} \left[ \dot{\psi_\pi} \dot{\psi_\pi} \right] + \sum_{n = 1}^{\infty} (-1)^n\text{Tr}\left[\dot{\psi_\pi}(n) \dot{\psi_\pi}\right] \right) \nonumber \\
    &= \frac{1}{A_\infty^\pi (0)2^L} \sum_{i,j} |\langle i| \Tilde{V} |\Tilde{j}\rangle|^2  \pi \delta_F\left(\epsilon_i -\epsilon_j + \frac{\pi}{T} \right).
    \label{Eq: Fermi's Golden Rule PiMode}
\end{align}
In contrast to the zero mode, the $(-1)^n$ prefactor appears in the first line of \eqref{Eq: Fermi's Golden Rule PiMode}, and is absorbed into the delta function by the inclusion of the $\pi/T$ factor in the argument.  The notation in the second line is slightly different from that for the zero mode, and is as follows: $\Tilde{V} = V - \psi_\pi V \psi_\pi $ and $|\Tilde{j} \rangle = \psi_\pi |j\rangle$.

\begin{figure*}[t]
    \centering    \includegraphics[width=0.4\linewidth]{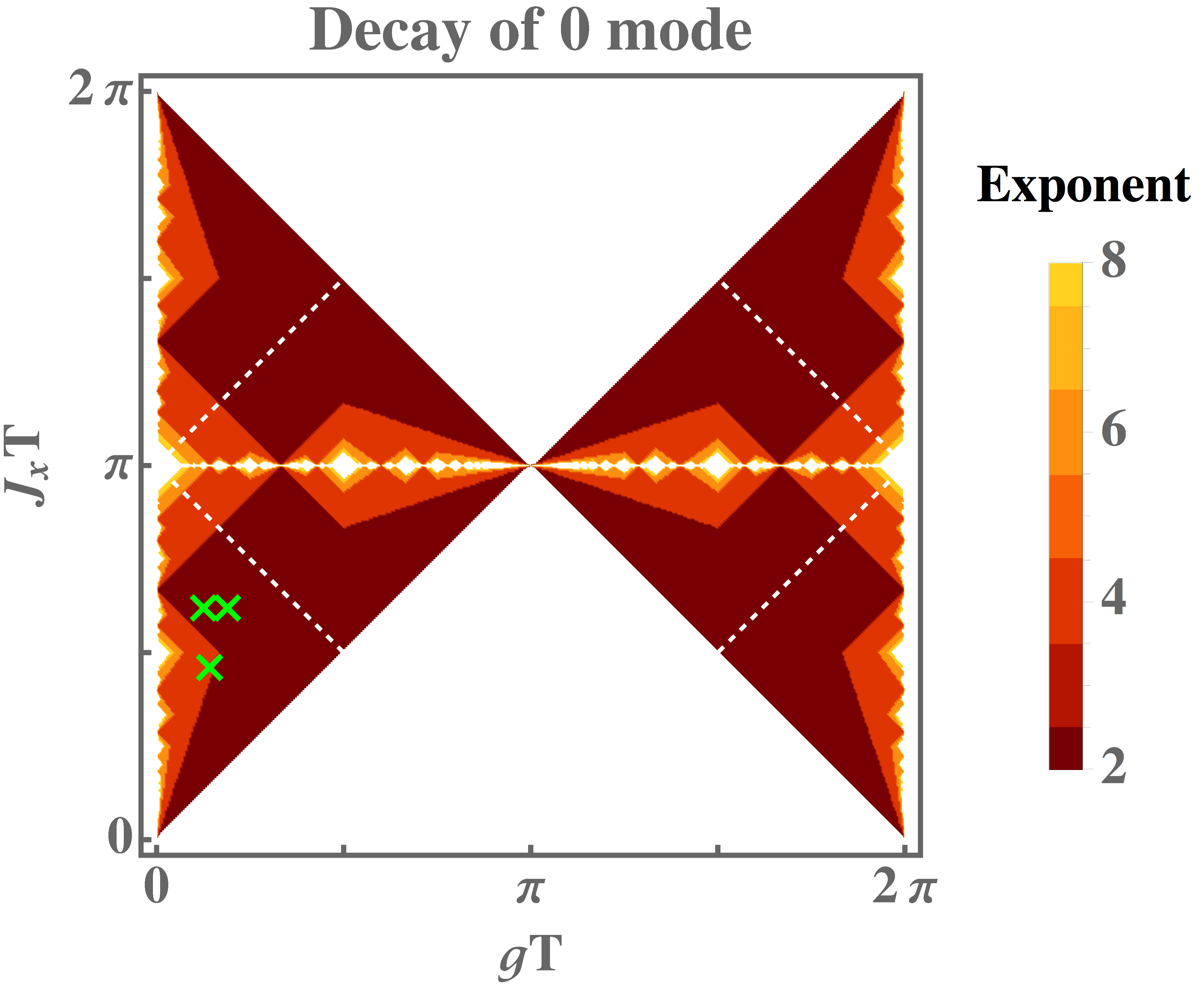} \ \includegraphics[width=0.4\linewidth]{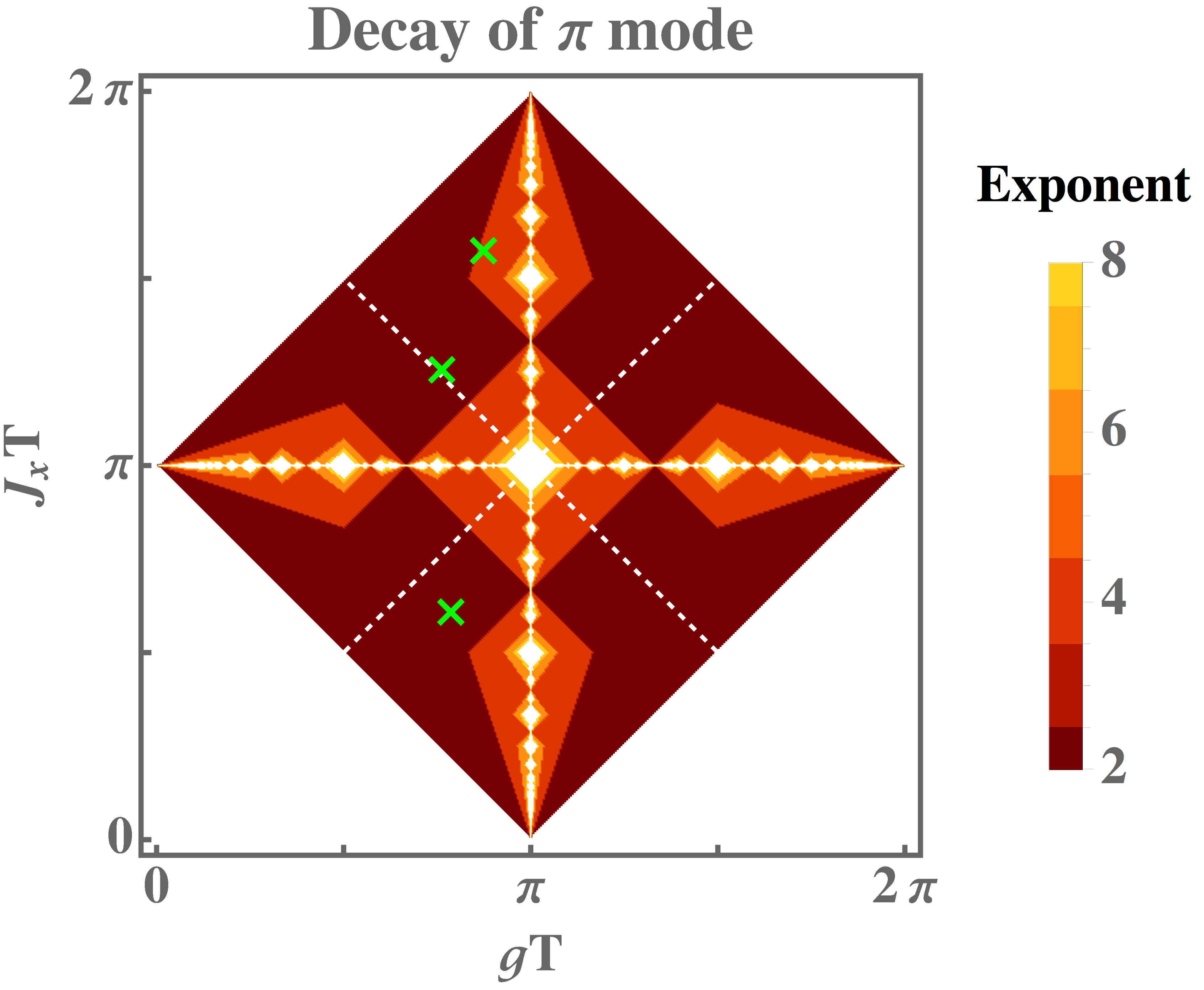}
    \caption{Plot of the exponent $m$ characterizing the decay of the zero mode (left panel) and the $\pi$ mode (right panel) by bulk scattering processes. The decay rate is predicted to be proportional to $J_z^m$ for $J_z \to 0$ where the exponent $m$ is encoded in the color of the plot.
The exponent $m=2 n$ is determined from the condition whether the energy of the mode is part of the $(4 n-1)$-particle  continuum, see Fig.~\ref{Fig:continua} and text. Only those regions in parameter space are shown where the corresponding mode exists. The dashed white lines show the boundary of the existence for $\pi$ mode (left panel) and 0 mode (right panel), and we focus on the region where only one kind of edge mode is allowed. The crosses denote parameter values investigated in Figs.~\ref{Fig: 0Mode Autocorrelation} and \ref{Fig: PiMode Autocorrelation}.
    \label{Fig:decayExponent}}
\end{figure*}

The decay rates derived above for zero and $\pi$ modes are only finite when the resonant conditions $\epsilon_i-\epsilon_j=0$ and $\epsilon_i-\epsilon_j-\frac{\pi}{T}=0$, respectively, are fulfilled (modulo $2 \pi/T$). Our perturbation is $V = J_z H_{zz}/2$ and after a Jordan Wigner transformation, this can be written as a four Majorana fermion interaction term, $\sigma_i^z \sigma_{i+1}^z = - a_{2i-1}a_{2i}a_{2i+1}a_{2i+2}$, where $a_i$ are the Majorana operators. 
As a next step one can express those in terms of the edge-state Majorana $\psi_0$ and $\psi_\pi$ and bulk  operators $d_k$ and $d^\dagger_k$.
The matrix element $|\langle i| \Tilde{V} |\Tilde{j}\rangle|$ in FGR survives only when the perturbation $V$ involves an edge mode such that $\Tilde{V}$ is non-zero.
To affect the edge mode, one of the operators has to be $\psi_\alpha$ with $\alpha=0,\pi$, the others can be bulk modes. Thus in the thermodynamic limit a decay of $\psi_0$ in second order perturbation theory is obtained if one finds a solution of the following equation
\begin{align}
0=\epsilon_{k_1}\pm \epsilon_{k_2}\pm \epsilon_{k_3} \qquad(\text{mod}\ 2 \pi/T).
 \end{align}
We therefore expect a non-zero decay rate of the $\psi_0$ mode at second order in $J_z$ only if the three particle-continuum, defined by the sums and differences of three bulk energies $\epsilon_k$, contains the energy $0$.
Fig.~\ref{Fig:continua} shows different $n$-particle continua as a function of $g T$ and for a fixed value of $J_x T=1.5$.

Similarly, the $\pi$ mode will decay if one finds a solution for 
 \begin{align}
\pi/T=\epsilon_{k_1}\pm \epsilon_{k_2}\pm \epsilon_{k_3} \qquad(\text{mod}\ 2 \pi/T).
 \end{align}
Thus one has to check whether the three- particle continuum contains the quasi-energy $\pi/T$.
If both $\pi$ and zero modes are present, scattering processes proportional to $\psi_0 \psi_\pi$ and two bulk operators are possible, leading to the condition
  \begin{align}
\pi/T=\epsilon_{k_1}\pm \epsilon_{k_2} \qquad(\text{mod}\ 2 \pi/T).
 \end{align}
This process is activated if the two-particle continuum includes the energy $\pi/T$.
In this paper we discuss the decay of isolated $0,\pi$ modes, leaving the
discussion of the decay when both modes are present, to a later publication.

The conditions discussed above, easily generalize to higher-order scattering processes. Matrix elements arising to order $J_z^{2 n}$ involve maximally $4 n$ Majorana fermions and thus maximally $4 n -1$ bulk modes. One therefore obtains a contribution to the decay rate of $\psi_0$ to order $J_z^{2 n}$ if a solution exists for
 \begin{align}
0=\sum_{i=1}^{4n-1} (\pm \epsilon_{k_i})  \qquad(\text{mod}\ 2 \pi/T),
\label{Eq: 0-mode energy conservation}
 \end{align}
or, equivalently, if the energy $0$ is part of the $(4 n-1)$-particle continuum of the bulk states. The condition of \eqref{Eq: 0-mode energy conservation} strongly restricts the phase space (i.e., the subset of allowed $k_i$ values) available for scattering in a $J_z^{2 n}$-scattering process.
Similarly, the decay of the $\pi$ mode is triggered for  
\begin{align}
\pi/T=\sum_{i=1}^{4n-1} (\pm \epsilon_{k_i})  \qquad(\text{mod}\ 2 \pi/T).
\label{Eq: Pi-mode energy conservation}
 \end{align}
If both $0,\pi$ modes are present, there is a further decay channel arising from the $(4n-2)$-particle continuum
 \begin{align}
\pi/T=\sum_{i=1}^{4n-2} (\pm \epsilon_{k_i}) \qquad(\text{mod}\ 2 \pi/T),
 \end{align}
a regime we plan to explore in future work.
The knowledge of the maxima and minima of the bulk dispersion $\epsilon_k$ of the integrable system are sufficient to construct analytically the $m$-particle continuum (see Appendix \ref{sec:D}). As  shown in Fig.~\ref{Fig:continua}, the larger the number of excitations, the larger is the range of quasi-energies which can couple to the modes. 

Thus, without any further calculation, one can determine the leading power  $m$ in the decay rate, $\Gamma \sim J_z^m$, of the zero or $\pi$ modes. For a given set of parameters, one has to determine the smallest value of $n$ which leads to a finite $4 n-1$-particle continuum at either the quasi-energy zero or $\pi$, see Fig.~\ref{Fig:continua}. This determines the exponent $m=2n$ in the $J_z \to 0$ limit.
This exponent is shown in Fig.~\ref{Fig:decayExponent},  in the region where the edge modes exist, and accounting for $m$ upto $8$. For a large part of the phase diagram one obtains $\Gamma \sim J_z^2$, but there are also sizable regions in the phase diagram, where larger exponents are obtained, resulting in a much longer lifetime of the edge modes. The exponents $m$ get larger and larger upon approaching the lines $g T=0, \pi$ or $J_x T=0, \pi$, which is explained by the fact that the quasi-particle bands become exactly flat in this limit, see \eqref{Eq: Bulk Dispersion}. For such exactly flat bands, $\epsilon_k=\epsilon_c={\rm constant}$, one cannot find any solution for \eqref{Eq: 0-mode energy conservation} and \eqref{Eq: Pi-mode energy conservation} (with the exception of points where $\epsilon_c/(2 \pi/T)$ is a rational number). Thus, the decay rate $\Gamma$ of edge modes becomes smaller than any power law in this limit for $J_z \to 0$.

It is worth mentioning that the  FGR formula is valid for any perturbation regardless of the symmetry. For example, a perturbation in the form of a $y$-direction transverse-field breaks $Z_2$ symmetry. However, in second order, such a perturbation already involves many-Majorana scattering processes since $\sigma_l^y$ is a string of Majoranas, $\sigma_l^y = (-ia_1a_2)\ldots(-ia_{2l-3}a_{2l-2})a_{2l}$. Here we consider the $Z_2$ symmetric perturbation $V = J_z H_{zz}/2$ such that the perturbing term has only 4 Majoranas. This allows for a simpler physical picture for describing different decay channels.

In the following section, we present numerical results to support these ideas.

\section{Results and discussion} \label{sec:Results}
\subsection{Almost strong zero mode}
\begin{figure*}[t]
    \centering      \includegraphics[width=0.32\textwidth]{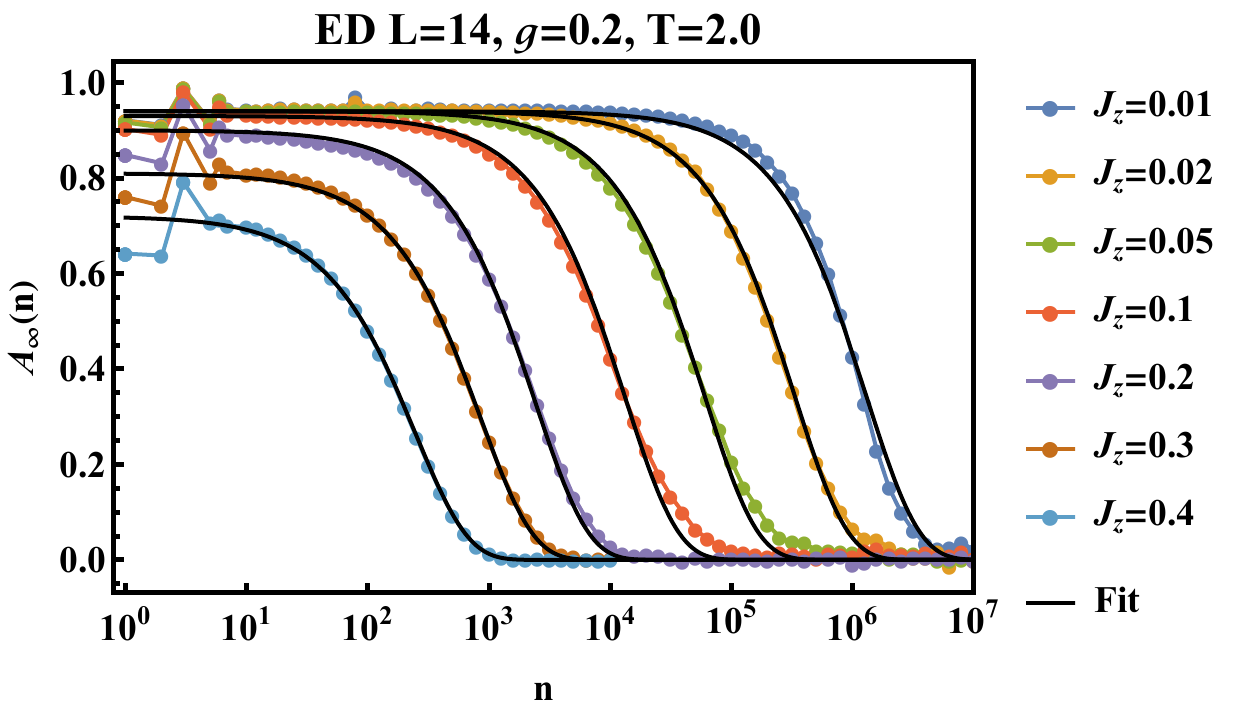} \includegraphics[width=0.32\textwidth]{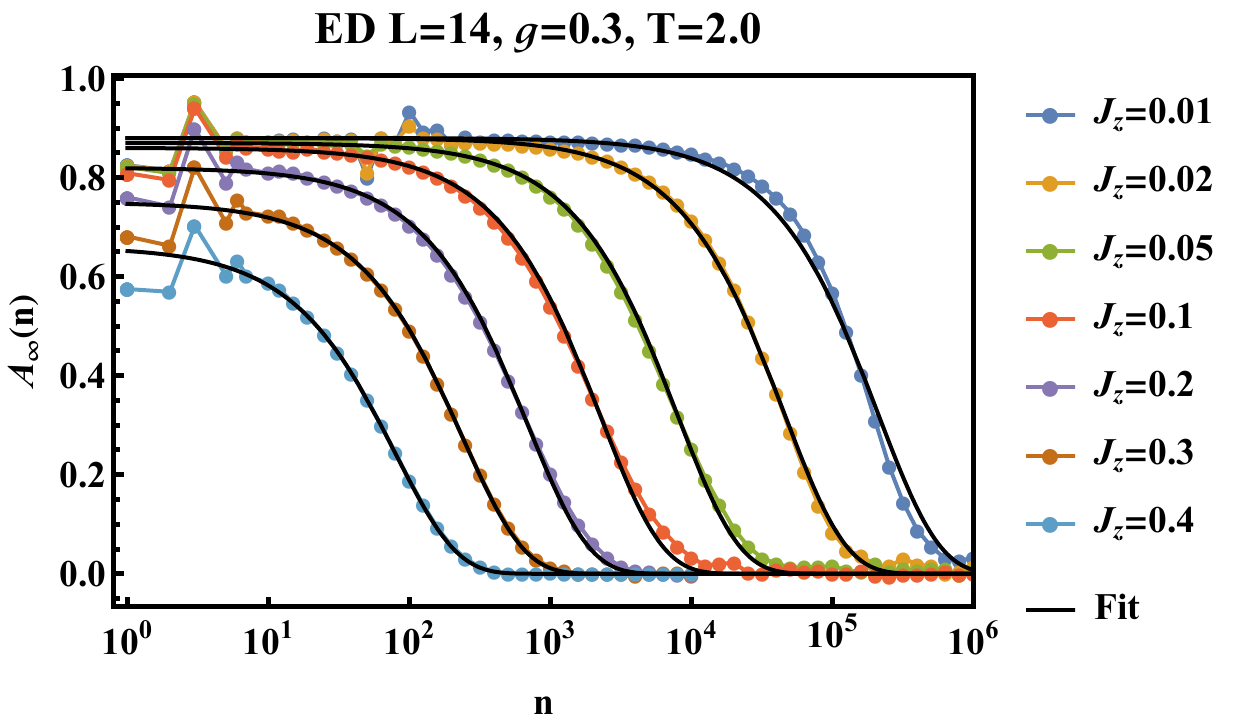}   \includegraphics[width=0.32\textwidth]{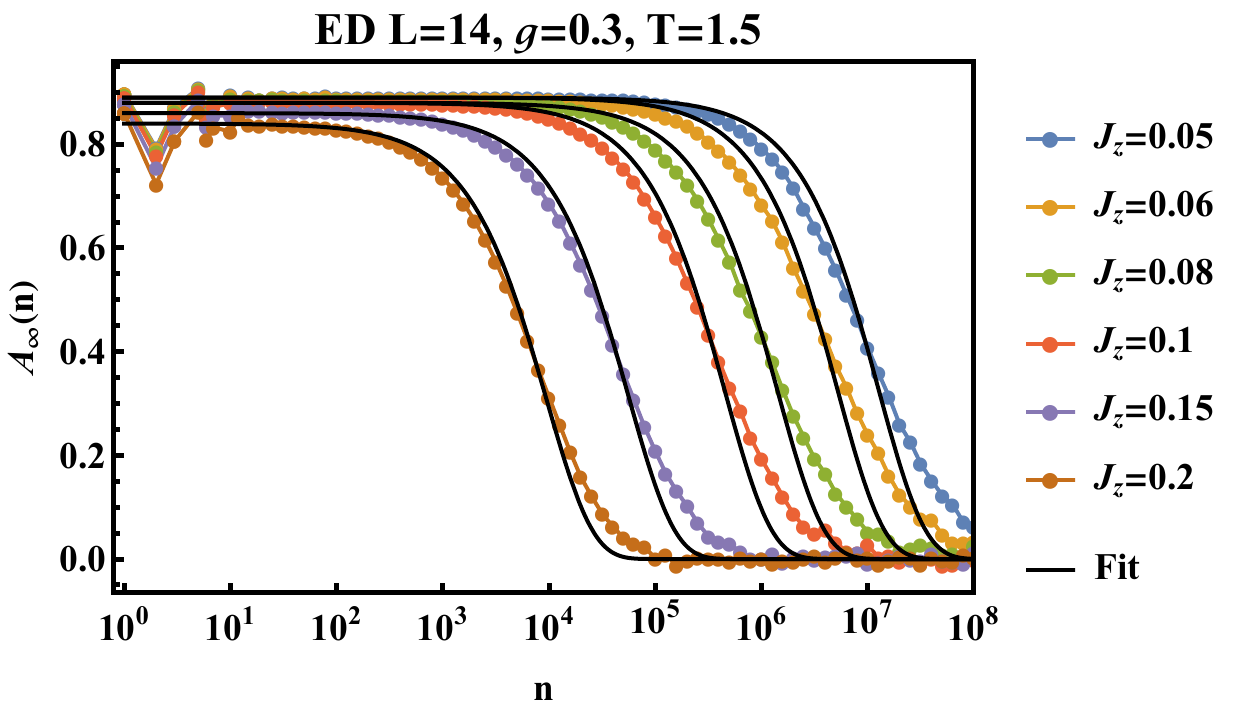} \includegraphics[width=0.32\textwidth]{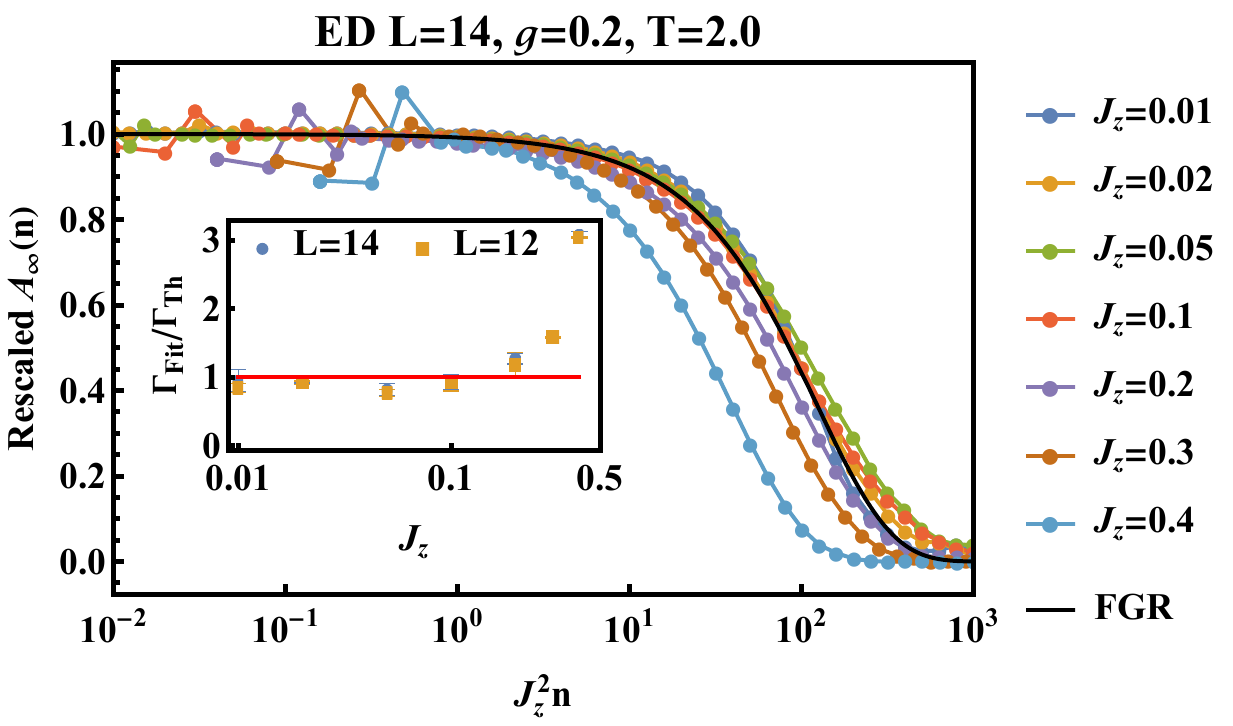} \includegraphics[width=0.32\textwidth]{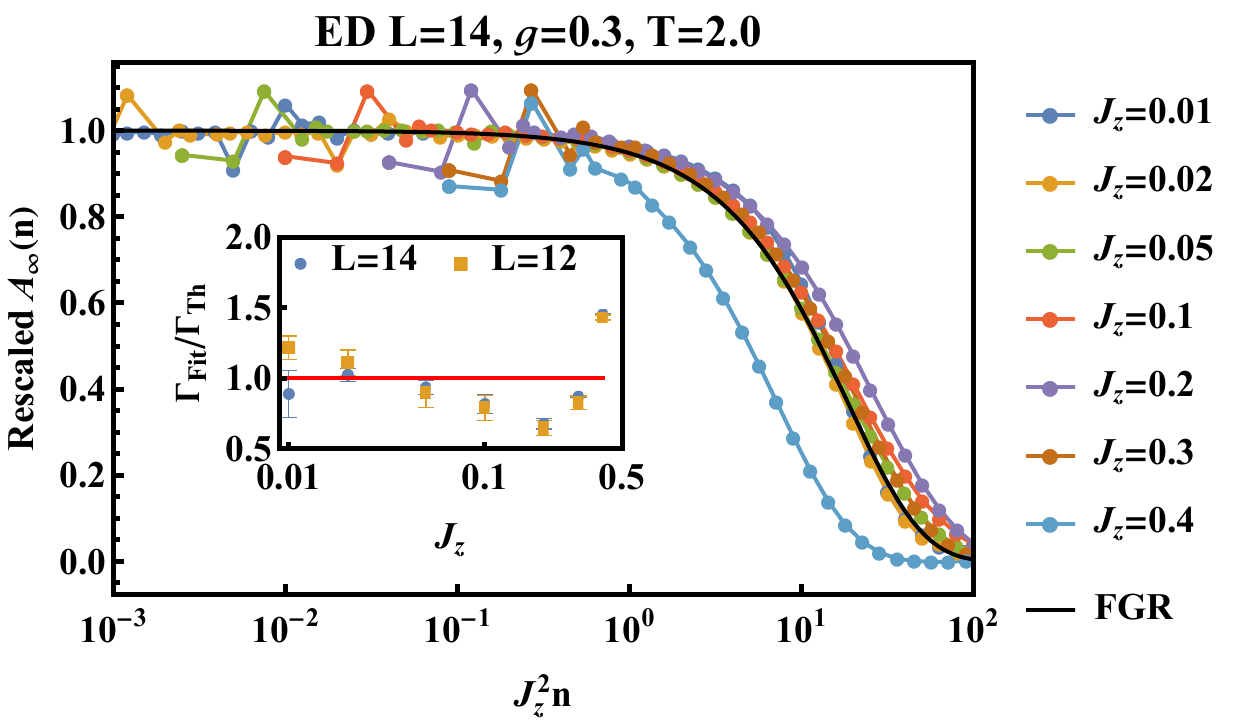} \includegraphics[width=0.32\textwidth]{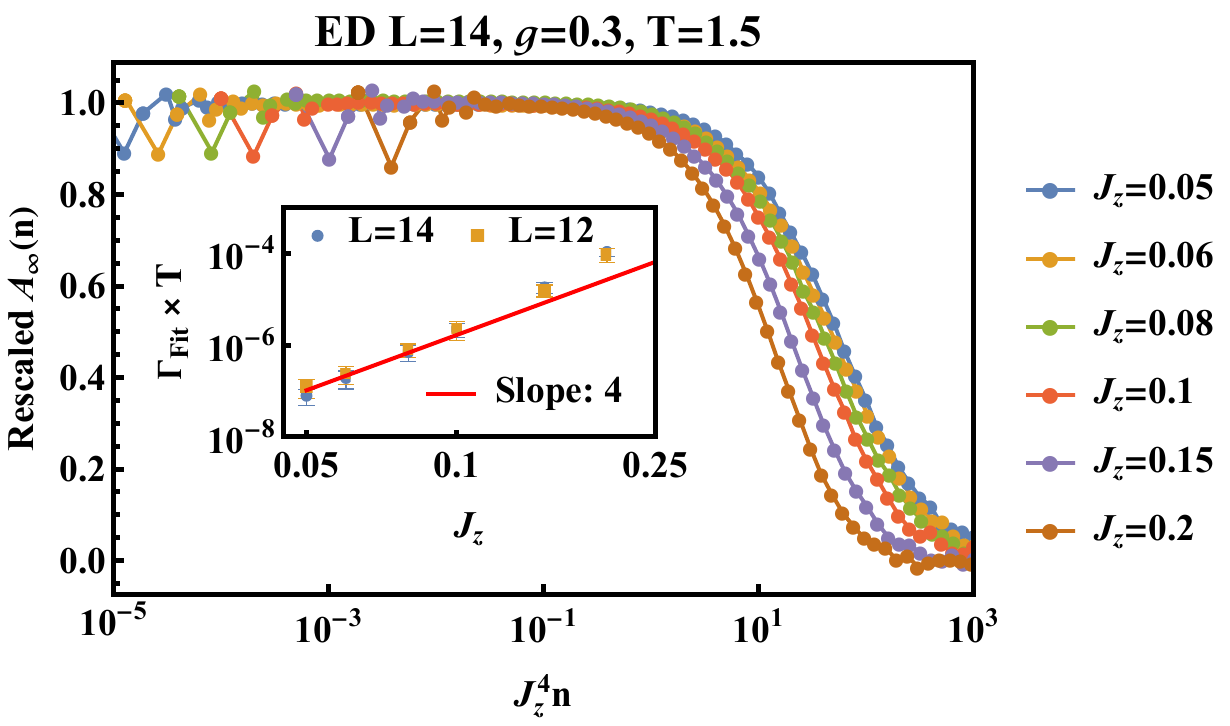}   
    \caption{Almost strong zero mode.
    Top panels: The infinite temperature autocorrelation of $\sigma_1^x$ for $L=14$ with $g = 0.2$, $T = 2.0$ (left), $g = 0.3$, $T = 2.0$ (middle) and $g = 0.3$, $T = 1.5$ (right) for different strengths of the integrability breaking term $J_z$. As $J_z$ decreases, the lifetime increases. Each data set is fitted with an exponential function $C \exp(-NT\Gamma_{\text{fit}})$ where the decay rate is determined from the average $\Gamma_{\text{fit}} = (\Gamma_{80\%} + \Gamma_{20\%})/2$ where $\Gamma_{80\%(20\%)}$ is the inverse time at which the correlator is $0.8(0.2)C$. Bottom panels: The autocorrelation function is rescaled to be $1$ in the quasi-stable region, and the time is rescaled to $J_z^{2}n$ ($J_z^{4}n$) for the second (fourth) order perturbation process. For $g =0.2$, $T=2.0$ (left) and $g =0.2$, $T=3.0$ (middle), the rescaled autocorrelation function  approaches the FGR theoretical values $\Gamma_\text{Th} T$ which equals $0.0082 J_z^2$ (left) and $0.054 J_z^2$ (middle) for small $J_z$. The inset shows the ratio between the numerically fitted decay rate $\Gamma_{\text{Fit}}$ and  the theoretical decay rate $\Gamma_\text{Th}$ for $L=12,14$. The horizontal red line marks  $\Gamma_{\text{Fit}}/\Gamma_\text{Th} = 1$ and the error bar of each data point shows the range between $\Gamma_{80\%}$ and $\Gamma_{20\%}$. The numerical results validate second order perturbation in $J_z$. For $g=0.3$, $T=1.5$ (right), the rescaled autocorrelation function and the inset that plots $\Gamma_{\text{Fit}} T$ vs. $J_z$ on a  log-log scale, are consistent with $\Gamma_{\text{Fit}} \propto J_z^4$ for small $J_z$.}
    \label{Fig: 0Mode Autocorrelation}
\end{figure*}

We first focus on the almost strong zero mode and how it is influenced by second order processes. Two cases are studied, $g = 0.2,0.3$ with $T=2.0$ where the system possesses only a zero mode. They correspond to the top two crosses in the left panel of Fig.~\ref{Fig:decayExponent}.
The decay rate is dominated by second order perturbation as the 3-particle continuum is closed at zero quasi-energy. 
The autocorrelation function of $\sigma_1^x$ are computed from exact diagonalization, with the results for $L = 14$ presented in top panels (left and middle) of Fig.~\ref{Fig: 0Mode Autocorrelation}. The rescaled plots are shown in the corresponding bottom panels. The rescaled autocorrelation functions approach the FGR prediction for small $J_z$. In addition, the numerically fitted decay rates, shown for two different system sizes, are in agreement with the perturbative result \eqref{Eq: Fermi's Golden Rule 0Mode}, see inset of Fig.~\ref{Fig: 0Mode Autocorrelation}.

To go beyond the region where second order perturbation theory is valid,
we consider a shorter period $T = 1.5$ with $g = 0.3$. This corresponds to the lower single cross in the left panel of Fig.~\ref{Fig:decayExponent}. Here, the system only possesses a zero mode but the resonance condition is satisfied by fourth order perturbation instead of second order perturbation theory.
Notice that the third order perturbation cannot be the leading order here because the decay rate is a positive number, and it should stay positive when $J_z \rightarrow -J_z$. Therefore, the leading order contribution can only be an even order perturbation. 

The autocorrelation function and the numerically fitted decay rate are presented in the right panels of Fig.~\ref{Fig: 0Mode Autocorrelation}. 
As we are probing a higher order perturbation process, the time-scales for a given value of $J_z$ become much longer, and
finite system size effects and deviation from exponential behavior (see below) become more apparent in comparison to the second order cases. In addition, the value of $J_z$ cannot be taken to be as small as in the second order region. The fitted decay rates and the plot of the autocorrelation functions as function of $J_z^4 n$ in the lower right panel of Fig.~\ref{Fig: 0Mode Autocorrelation} clearly support a decay rate proportional to $J_z^4$ in the small $J_z$ limit.

 Deviations from simple exponential behavior  appear both in regimes where the decay rates are proportional to $J_z^2$ and $J_z^4$. In the middle panel of Fig.~\ref{Fig: 0Mode Autocorrelation}, the autocorrelation function starts to clearly deviate from exponential decay for $J_z = 0.01$. Larger effects are seen in the right panel of Fig.~\ref{Fig: 0Mode Autocorrelation}, where decay rates are proportional to $J_z^4$. For all parameters where the curves do not follow a simple exponential decay in the long-time limit, we also observe finite size effects as shown in Appendix \ref{sec:E} where numerical results for $L=12$ and $14$ are compared. Mathematically, deviations from exponential behavior at the long time-scale $n T$, $n \gg 10^3$, reflects that the imaginary part of the self-energy of the Majorana mode depends on frequency in the small frequency scale $\sim 1/(n T)$. This arises because bulk quasi-energies are discrete in a finite size system. Nevertheless, as shown by the scaling plots, finite system size effects do not spoil the key features of the perturbative processes in the parameter region we have probed. Hence, the numerical results fully support the perturbative argument.

\subsection{Almost strong $\pi$ mode}
\begin{figure*}[t]
    \centering      \includegraphics[width=0.32\textwidth]{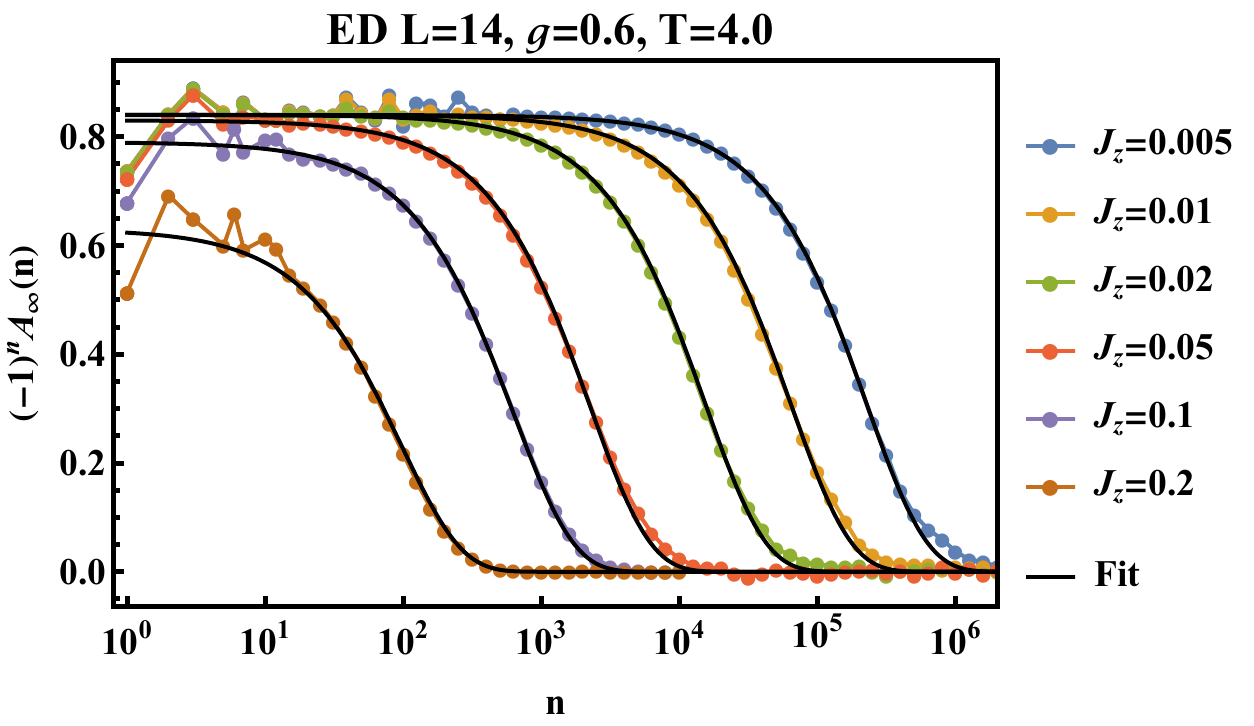} \includegraphics[width=0.32\textwidth]{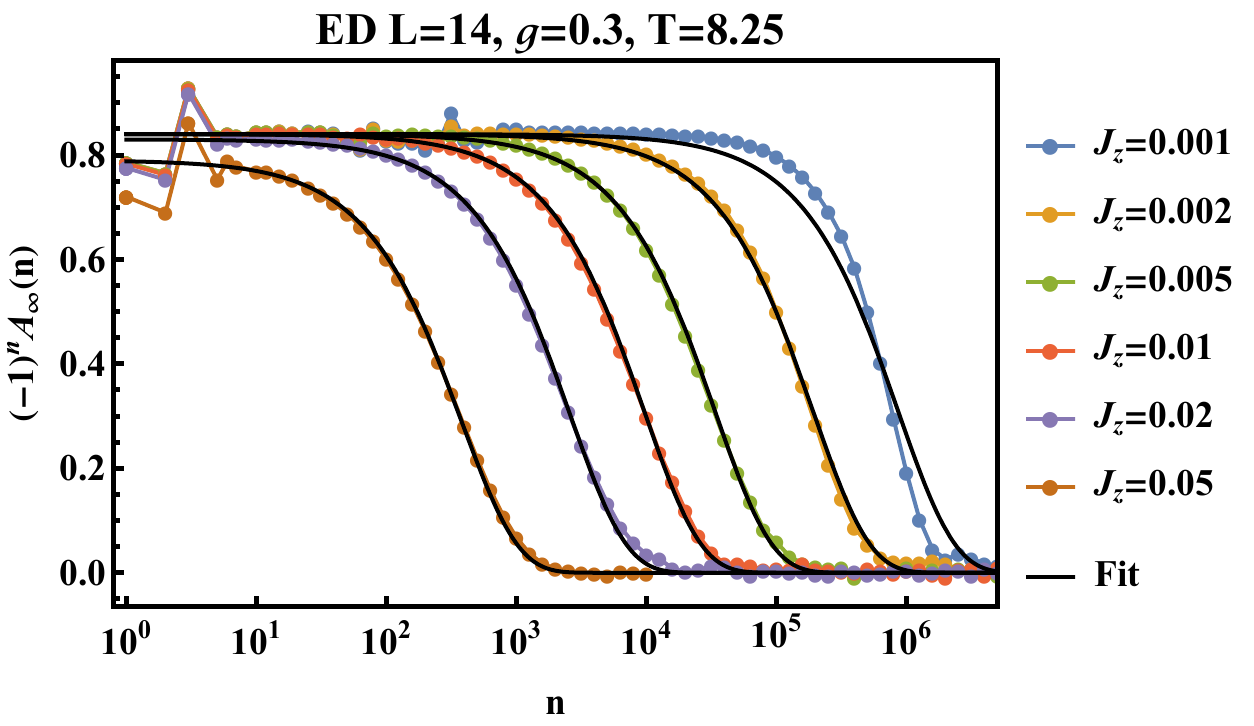}   \includegraphics[width=0.32\textwidth]{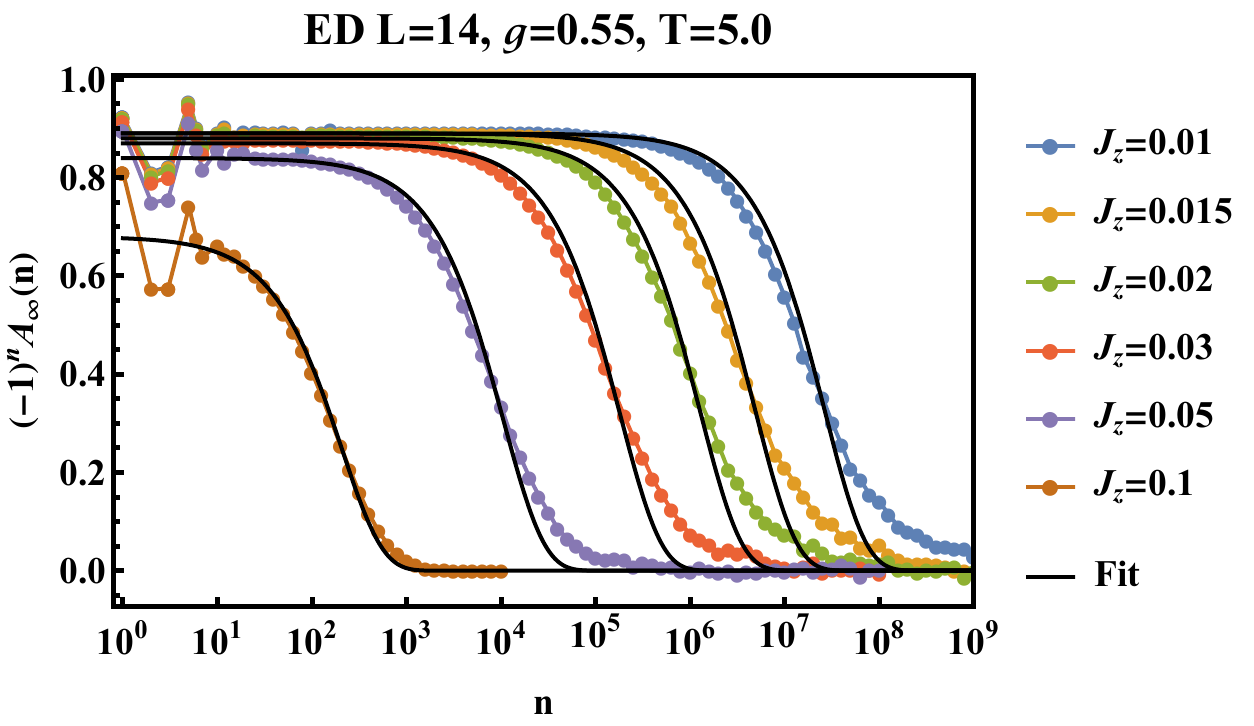} \includegraphics[width=0.32\textwidth]{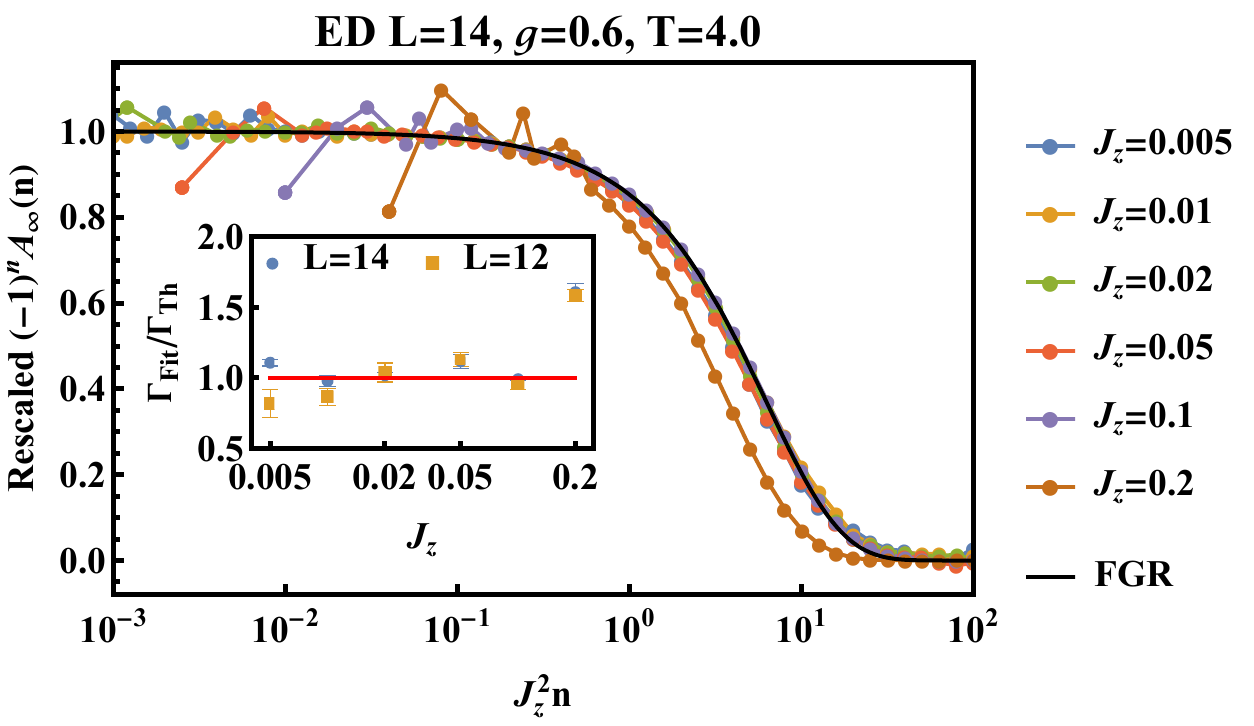} \includegraphics[width=0.32\textwidth]{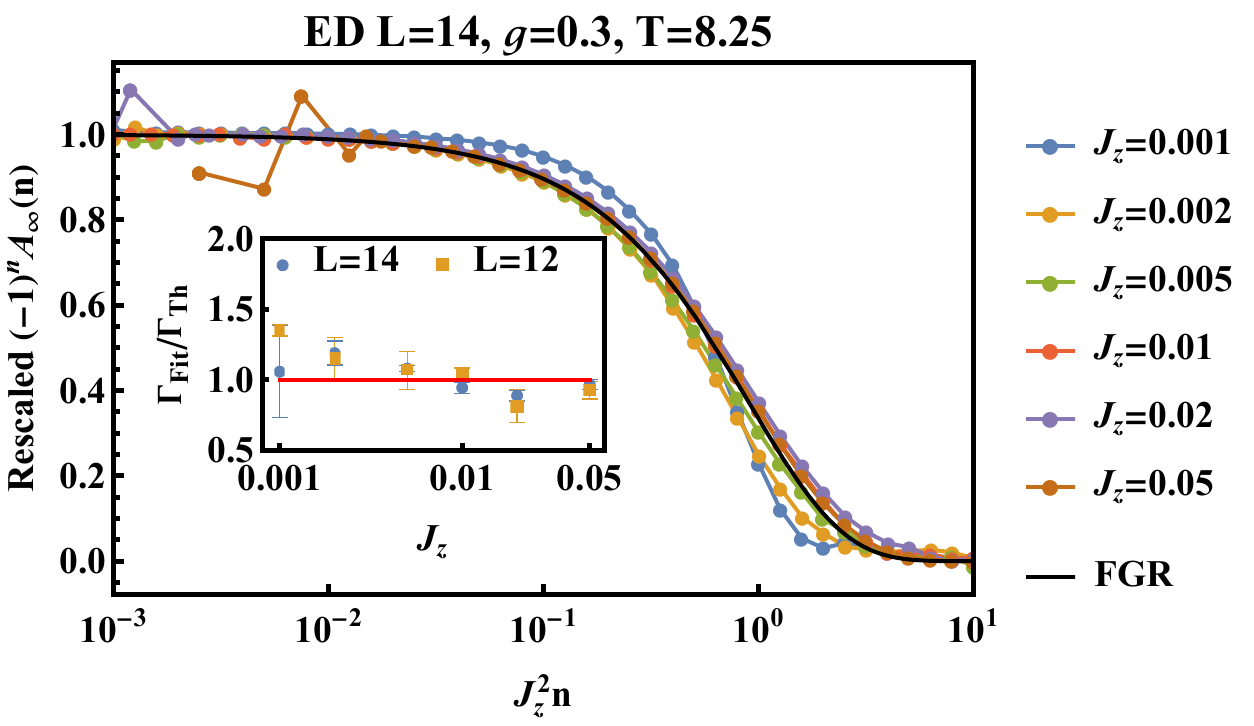} \includegraphics[width=0.32\textwidth]{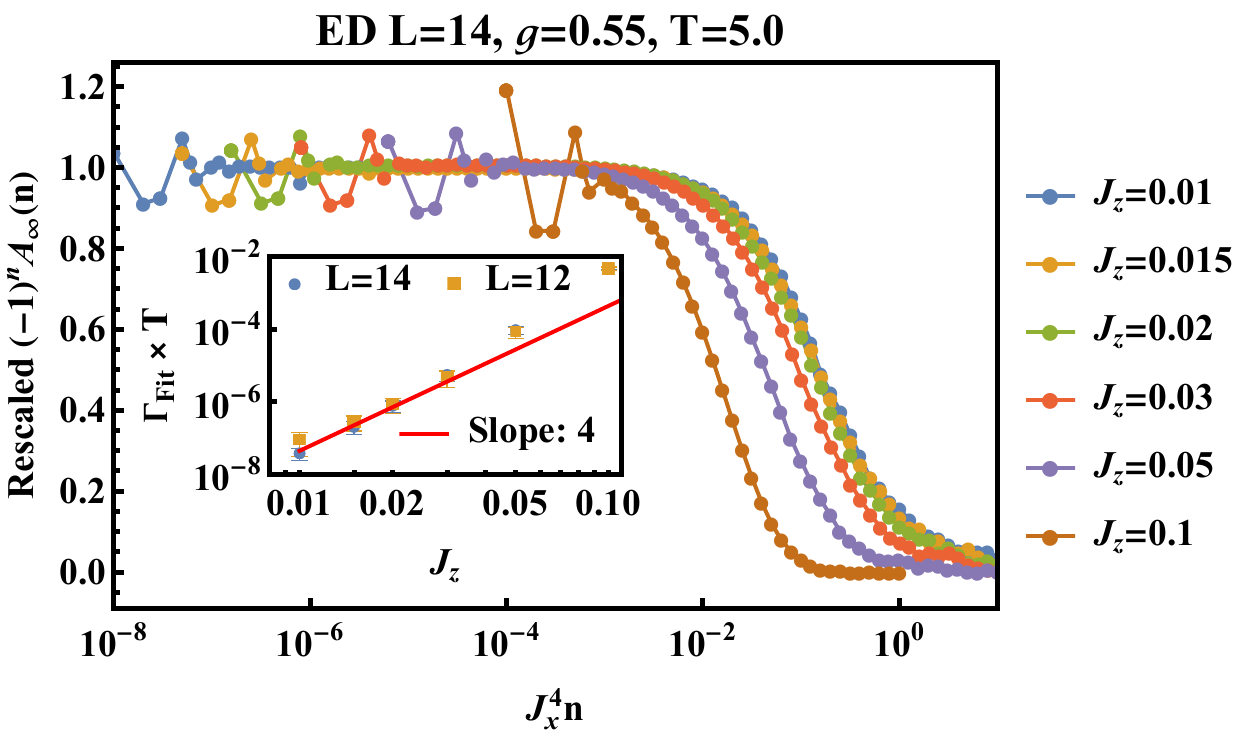}   
    \caption{Almost strong $\pi$ mode. 
    Top panels: The autocorrelation function of $\sigma^x_1$ for $L=14$ with $g = 0.6$, $T = 4.0$ (left), $g = 0.3$, $T = 8.25$ (middle) and $g = 0.55$, $T = 5.0$ (right) for different $J_z$. Here we multiply the autocorrelation function by $(-1)^n$ so that the time-evolution is smooth, like a zero mode. In addition, we fit the time-evolution to  an exponential decay in the same manner as was done for the zero mode case discussed in Fig.~\ref{Fig: 0Mode Autocorrelation}. Bottom panels: The corresponding rescaled autocorrelation function vs rescaled time $J_z^{2}n$ ($J_z^{4}n$) for the second (fourth) order perturbation process. For $g =0.6$, $T=4.0$ (left) and $g =0.3$, $T=8.25$ (middle), the rescaled autocorrelation function approaches the FGR theoretical values $\Gamma_\text{Th} T$ which equal $0.16 J_z^2$ (left) and $1.1 J_z^2$ (middle) for small $J_z$. The inset shows the ratio between the numerically fitted decay rate $\Gamma_{\text{Fit}}$ and  the theoretical decay rate $\Gamma_\text{Th}$ for $L=12,14$. The horizontal red line marks  $\Gamma_{\text{Fit}}/\Gamma_\text{Th} = 1$ and the error bar of each data point shows the range between $\Gamma_{80\%}$ and $\Gamma_{20\%}$. The numerical results validate second order perturbation in $J_z$. For $g=0.3$, $T=1.5$ (right), the rescaled autocorrelation function and the inset showing $\Gamma_{\text{Fit}} T$ vs. $J_z$ on a  log-log scale, are consistent with $\Gamma_{\text{Fit}} \propto J_z^4$ for small $J_z$.}
    \label{Fig: PiMode Autocorrelation}
\end{figure*}

Now, we turn to the decay of the $\pi$ mode.  We first focus on two different periods, $T=4.0$ and $T=8.25$, that allow the existence of a $\pi$ mode for a range of $g$.
We study two cases, $g=0.6, T=4.0$ and $g=0.3, T=8.25$ corresponding to the middle and bottom crosses in the right panel of Fig.~\ref{Fig:decayExponent}. The decay is due to resonances that are second order in $J_z$. 
The autocorrelation function is presented in the top panels (left and middle) of Fig.~\ref{Fig: PiMode Autocorrelation}. In the corresponding bottom panels in Fig.~\ref{Fig: PiMode Autocorrelation}, the rescaled autocorrelation function is plotted and shown to approach the FGR prediction for small $J_z$. The numerically fitted decay rates are plotted in the corresponding insets of Fig.~\ref{Fig: PiMode Autocorrelation} and are shown to be consistent with the FGR result \eqref{Eq: Fermi's Golden Rule PiMode}.

To explore resonances arising from higher order processes, a different period $T = 5.0$ is considered with $g =0.55$. This corresponds to the top cross in the right panel in Fig.~\ref{Fig:decayExponent},
where the system only possesses a $\pi$ mode but the resonance condition is matched by fourth order perturbation  theory instead of second order perturbation theory.
The autocorrelation function and the numerically fitted decay rate are presented in the right panels of Fig.~\ref{Fig: PiMode Autocorrelation}. As with the zero mode case, the finite system size effect is apparent at $L=14$ for higher order processes, making the range of $J_z$ that can be probed, more limited. Despite the deviation from exponential decay of the autocorrelation function, we can still estimate the decay rate by the same fitting method as before since the decay rates for $L=12,14$ overlap for $J_z$ values as small as $J_z = 0.01$. The rescaled autocorrelation function plotted in the bottom right panel tends to saturate for small $J_z$. Qualitatively, the scaling of the decay rate shows a behavior consistent with $J_z^4$ for small $J_z$, in agreement with the resonance conditions deduced from studying the many-particle continuum.

The finite system size effects are visible in both second and fourth order dominated processes. When second order processes are dominant, the finite system size effect is severe for $g=0.3$, $T=8.25$, for the small $J_z$ value of $J_z = 0.001$ (middle panels). Here, the autocorrelation function deviates considerably from an exponential decay, and fails to collapse to the FGR prediction. This can also be observed in the increasing error bars in the fitted decay rate. On the contrary, the finite system size effect is not essential for $g=0.6$, $T=4.0$ (left panels) as the rescaled autocorrelation functions collapse well around the FGR prediction for small $J_z$. When fourth order processes dominate (right panels), as with the zero mode case, the finite size effects become more severe, with  the autocorrelation functions following a slower decay than exponential for small $J_z$. A more detailed discussion of finite size effects is presented in  Appendix \ref{sec:E}.

Finally, we would like to stress the connection between this work and prethermalization \cite{Nayak17,else2017prethermalfloquet}.
If in a Floquet system the driving frequency $\omega$ is much larger than all other energy scales, then the heating rate, which drives the system to infinite temperature, is exponentially small \cite{Abanin15,abanin2017rigorous}. Here, we investigate a somewhat different question: we assume that the system has already reached infinite temperature, and we explore under this condition, the stability of topological boundary modes. We are also not focusing on the limit where the frequency is much larger than the relevant bandwidth. Nevertheless, the physics which leads to exponentially small decay rates is actually fully consistent with our approach. In Fig.~\ref{Fig:decayExponent}, this is encoded in the fact that the exponent $n$ gets larger and larger when one approaches parameters where the Floquet bands become flat ($J_x T=0,\pi$ or $g T=0,\pi$). A power-law decay rate $\Gamma \sim J_z^{2n} =e^{-2n \ln(1/J_z)}$ is equivalent to an exponential suppression (with logarithmic corrections), $\Gamma \sim e^{-c/W}$, provided that $n \propto 1/W$, where $W$ is the bandwidth of the relevant band. Such a behavior follows from \eqref{Eq: 0-mode energy conservation} and \eqref{Eq: Pi-mode energy conservation} when one considers first the limit of vanishing $J_z$ and afterwards the limit of a vanishing bandwidth (the two limits do not commute). 

To see this more clearly, note that for a decay rate $\propto J_z^{2n} $ of the zero mode, one needs to satisfy
the following modulo $2\pi$: $n_1\epsilon_{\rm max} = n_2\epsilon_{\rm min}, n_1+n_2=4n-1$ (see Appendix \ref{sec:D}). This implies,
$n_1 + n_2 = (n_2-n_1) (\epsilon_{\rm max}+\epsilon_{\rm min})/(\epsilon_{\rm max}-\epsilon_{\rm min})$.
Thus, in the limit of a narrow bandwidth $J_z \ll (\epsilon_{\rm max}-\epsilon_{\rm min})/(\epsilon_{\rm max}+\epsilon_{\rm min})\ll 1$, we expect that the decay rate is approximately given by $J_z^{2n} \sim e^{-c/W}$, where $W=\epsilon_{\rm max}-\epsilon_{\rm min}$ and $c$ depends logarithmically on $J_z$. This argument does not however take into account $n$-dependent combinatorial prefactors in the decay rate, which can lead to extra logarithmic corrections, see Refs. \cite{Abanin15,abanin2017rigorous}.

\section{Conclusions} \label{sec:Conc}

While periodically driven many-particle systems tend to heat up to infinite temperatures, almost strong modes can, nevertheless, have very long life times. Besides the size of integrability breaking terms, here the decisive factor is the phase space available for scattering. The conservation of quasi-energies governs which states are available for scattering and thus controls in which order of perturbation theory one can obtain finite decay rates. Our study tests this physics numerically in the perhaps most simple setting of a Floquet version of the one-dimensional Ising model which can host two types of almost strong modes, zero and $\pi$ modes. A major advantage of the model is that it can be simulated in a  straightforward way on quantum computers \cite{Abanin-Aleiner22}, even if present-day devices are too noisy to explore the extremely long time scales relevant in our study. 

Alternatively, our study can be viewed as an investigation of the stability of Majorana modes in static or periodically driven \cite{Matthies2022} one-dimensional topological superconductors with respect to quasi-particle poisoning. 
While nominally described by the same type of Hamiltonian, the edge modes in the superconducting realization are more stable with respect to noise, with a low-temperature environment strongly reducing scattering in comparison to our infinite-temperature calculation. In the Floquet case, the precise preparation protocol could also play a decisive role for the stability of the system \cite{Matthies2022}. Although we focus on a one-dimensional system in this work, the Floquet FGR is generic and can be applied to study systems in higher dimension, e.g., two-dimensional higher-order Floquet topological superconductors \cite{vu2021superconductors,ghosh2021floquet}.

To further explore the stability of almost strong modes and topological qubits, either in solid state realizations or in a quantum-computer, it will be interesting to extend our study to noisy environments and to systems where phonons provide on the one hand cooling, and on the other hand novel quasiparticle poisoning channels.

{\sl Acknowledgments:} 
This work was primarily supported by the US Department of Energy, Office of
Science, Basic Energy Sciences, under Award No.~DE-SC0010821 (HY, AM) and by the  German Research Foundation within
CRC183 (project number 277101999, subproject A01 (AR) and partially by the Mercator fellowship (AM)). HY acknowledges support of  the NYU IT High Performance Computing resources, services, and staff expertise.


\appendix
\section{Bulk dispersion relation in Floquet transverse-field Ising Model} \label{sec:A}
In this appendix, we present a detailed derivation of the bulk dispersion of the Floquet transverse-field Ising Model. The same techniques can also be applied in the continuous time case. The Floquet unitary is
\begin{align}
    &U_0 = U_z U_{xx},
\end{align}
where
\begin{align}
    &U_z = \exp\left(-ig \frac{T}{2} H_z\right);
    &U_{xx} = \exp\left(-i J_x  \frac{T}{2}H_{xx}\right).
\end{align}
$H_z$ and $H_{xx}$ are defined in \eqref{Eq: Ising Hamiltonian} describing transverse-field and Ising interactions respectively. The model can be mapped to a bilinear spinless Fermion model through the Jordan-Wigner transformation
\begin{align}
    &c_j^\dagger = \prod_{l < j}\sigma_l^z \sigma_j^-;
    &c_j = \prod_{l < j}\sigma_l^z \sigma_j^+,
\end{align}
where $\sigma_j^{\pm} = (\sigma_j^x \pm i\sigma_j^y)/2$. Therefore, the spin interactions can be expressed in terms of $c$ and $c^\dagger$ as follows
\begin{align}
    &\sigma_i^z = 1 - 2c_i^\dagger c_i,\\
    &\sigma_i^x \sigma_{i+1}^x = c_i^\dagger c_{i+1}^\dagger + c_i^\dagger c_{i+1} + c_{i+1}^\dagger c_i + c_{i+1} c_i.
\end{align}
We impose periodic boundary conditions and perform a Fourier transformation
\begin{align}
    &c_j = \frac{e^{i\pi/4}}{\sqrt{L}}\sum_k e^{ikj}c_k, &c_j^\dagger = \frac{e^{-i\pi/4}}{\sqrt{L}}\sum_k e^{-ikj}c_k^\dagger,
\end{align}
where $L$ is the system size. The spin Hamiltonians in $k$-space are
\begin{align}
    H_z &= \sum_{k>0} \biggl[c_k c_k^\dagger-c_k^\dagger c_k + c_{-k} c_{-k}^\dagger-c_{-k}^\dagger c_{-k}\biggr],\\
    H_{xx} &= 2\sum_{k>0} \biggl[\cos{k} (c_k^\dagger c_k - c_{-k} c_{-k}^\dagger) \nonumber\\
    & + \sin{k}(c_k^\dagger c_{-k}^\dagger + c_{-k} c_k)\biggr].
\end{align}
The Floquet unitary can now be written as a product of unitaries in $k$-space
\begin{align}
    U = \prod_{k>0} U_z^k U_{xx}^k,
\end{align}
where
\begin{align}
    U_z^k = \exp&\left[-i \frac{gT}{2} (c_k c_k^\dagger-c_k^\dagger c_k + c_{-k} c_{-k}^\dagger-c_{-k}^\dagger c_{-k})\right],\\
    U_{xx}^k = \exp& \left[-i J_x T \left(\cos{k} (c_k^\dagger c_k - c_{-k} c_{-k}^\dagger) \right.\right. \nonumber \\
      &+ \left.\left.\sin{k}(c_k^\dagger c_{-k}^\dagger + c_{-k} c_k) \right) \right].
\end{align}
It is useful to consider a $4\times 4$ matrix representation of $U_z^k$ and $U_{xx}^k$. We choose four orthogonal bases: $|0\rangle,\ |1\rangle = c_k^\dagger c_{-k}^\dagger|0\rangle,\ |2\rangle = c_k^\dagger|0\rangle$ and $ |3\rangle = c_{-k}^\dagger|0\rangle$. In this basis $ U_z^k$ and $U_{xx}^k$ are as follows
\begin{align}
    &U_z^k = 
    \begin{pmatrix}
    \exp(-igT) & 0 & 0 & 0 \\
    0 & \exp(igT) & 0 & 0\\
    0 & 0 & 1 & 0\\
    0 & 0 & 0 & 1
    \end{pmatrix},\\
    &U_{xx}^k =
    \begin{pmatrix}
        u_{xx} & 0_{2\times 2}\\
        0_{2\times 2} & \mathbb{I}_{2\times 2}
    \end{pmatrix},
\end{align}
where $u_{xx}$ is a 2 by 2 matrix
\begin{align}
    u_{xx} = \mathbb{I}_{2\times 2} \cos(J_x T)  -i \sin(J_x T) \left(\sigma_x \sin k  - \sigma_z\cos k  \right).
\end{align}
Finally, the multiplication of these two matrices leads to
\begin{align}
    U_k = U_z^k U_{xx}^k=
    \begin{pmatrix}
        \alpha & \beta & 0 & 0\\
        -\beta^* & \alpha^* & 0 & 0\\
        0 & 0 & 1 & 0\\
        0 & 0 & 0 & 1
    \end{pmatrix},
\end{align}
where
\begin{align}
    &\alpha = e^{-igT}[\cos(J_x T) + i \sin(J_x T) \cos{k}], \label{Eq: alpha}\\
    &\beta = -ie^{-igT}\sin(J_x T) \sin{k}.
\end{align}
Since $U_k U_k^\dagger = 1$, one obtains $|\alpha|^2 + |\beta|^2 = 1$. Focusing on the upper-left $2\times 2$ block,
the two eigenvalues $\exp(\pm i\epsilon_k T)$, are given by
\begin{align}
    \exp(\pm i\epsilon_k T) =  \text{Re}[\alpha] \pm i\sqrt{1-\text{Re}[\alpha]^2}.
    \label{Eq: epsilonk in alpha}
\end{align}
By examining the real part of \eqref{Eq: epsilonk in alpha} with \eqref{Eq: alpha}, one arrives at
\begin{align}
    \cos(\epsilon_k T) =  \cos(gT)\cos(J_x T) + \sin(gT)\sin(J_x T)\cos{k}.
\end{align}
Above is the bulk dispersion reported in \eqref{Eq: Bulk Dispersion}. Note that the quasi-energy band $\epsilon_k$ becomes exactly flat when either $J_x T=0, \pi$ or $g T= 0, \pi$.

The upper-left $2\times 2$ block can be rewritten in an exponential form by employing $|\alpha|^2 + |\beta|^2 = 1$ and \eqref{Eq: epsilonk in alpha}
\begin{align}
    \begin{pmatrix}
        \alpha & \beta \\
        -\beta^* & \alpha^*
    \end{pmatrix}
    =\exp\left[-i\frac{\epsilon_k T}{\sqrt{1-\text{Re}[\alpha]^2}}
    \begin{pmatrix}
        -\text{Im}[\alpha] & i\beta \\
        -i\beta^* & \text{Im}[\alpha]
    \end{pmatrix} \right].
\end{align}
Accordingly, the Floquet Hamiltonian for a given $k$-momentum is derived from $H_{F,k}^0 = i\ln(U_z^k U_{xx}^k)/T$. One obtains
\begin{align}
    H_{F,k}^0 &= \frac{\epsilon_k}{\sqrt{1-\text{Re}[\alpha]^2}}
    \begin{pmatrix}
        -\text{Im}[\alpha] & i\beta & 0 & 0\\
        -i\beta^* & \text{Im}[\alpha] & 0 & 0\\
        0 & 0 & 0 & 0\\
        0 & 0 & 0 & 0
    \end{pmatrix}.
\end{align}
Representing the above in terms of fermion bilinears, the full Floquet hamiltonian is
\begin{align}
    &H_F^0\nonumber\\ &= \sum_{k>0} \frac{\epsilon_k}{\sqrt{1-\text{Re}[\alpha]^2}}
    \begin{pmatrix}
        c_k^\dagger & c_{-k}
    \end{pmatrix}\!\!
    \begin{pmatrix}
        \text{Im}[\alpha] & -i\beta^*\\
        i\beta & -\text{Im}[\alpha]
    \end{pmatrix}  \!\!   \begin{pmatrix}
        c_k\\
        c_{-k}^\dagger
    \end{pmatrix}.
\end{align}
Diagonalizing the above via a Bogoliubov transformation leads to the Floquet Hamiltonian \eqref{Eq: Free Floquet Hamiltonian}
\begin{align}
    H_F^0 = \sum_{k>0} \epsilon_k 
    \begin{pmatrix}
        d_k^\dagger & d_{-k}
    \end{pmatrix}
    \begin{pmatrix}
        1 & 0\\
        0 & -1
    \end{pmatrix}
    \begin{pmatrix}
        d_k\\
        d_{-k}^\dagger
    \end{pmatrix},
\end{align}
where $(d_k^\dagger\  d_{-k}) = (c_k^\dagger\  c_{-k}) S$ with the transformation matrix $S$ given by
\begin{align}
    &S =\nonumber\\
    & 
    \begin{pmatrix}
        \frac{\sqrt{1-\text{Re}[\alpha]^2}+\text{Im}[\alpha]}{\sqrt{\left(\sqrt{1-\text{Re}[\alpha]^2}+\text{Im}[\alpha] \right)^2+|\beta|^2}} & \frac{i\beta^*}{\sqrt{\left(\sqrt{1-\text{Re}[\alpha]^2}+\text{Im}[\alpha] \right)^2+|\beta|^2}}\\
        \frac{i\beta}{\sqrt{\left(\sqrt{1-\text{Re}[\alpha]^2}+\text{Im}[\alpha] \right)^2+|\beta|^2}} & \frac{\sqrt{1-\text{Re}[\alpha]^2}+\text{Im}[\alpha]}{\sqrt{\left(\sqrt{1-\text{Re}[\alpha]^2}+\text{Im}[\alpha] \right)^2+|\beta|^2}}
    \end{pmatrix}.
\end{align}

\section{Zero mode and $\pi$ mode in Floquet transverse-field Ising Model} \label{sec:B}
In this appendix, we present analytic expressions for the zero and $\pi$ modes of the Floquet transverse-field Ising model. First, we introduce Majorana operators on odd and even sites following the convention that $l$ runs over $1$ to system size $L$.
\begin{align}
    &a_{2l-1} = \prod_{j=1}^{l-1}\sigma_j^z \sigma_l^x; & a_{2l} = \prod_{j=1}^{l-1}\sigma_j^z \sigma_l^y.
\end{align}
Next, we construct a generic operator as a linear combination of single Majorana operators, $\psi = \sum_{n} c_n a_n$. After one period of time evolution, the operator is still a superposition of single Majoranas
\begin{align}
    U_0^\dagger \psi U_0 &= K \psi,
\end{align}
where $\psi$ is $(c_0\ c_1 \cdots)^T$, a column vector representation of $\psi = \sum_{n} c_n a_n$. The corresponding $K$ in the same representation is
\begin{align}
    &K =\nonumber\\
    &
    \begin{pmatrix}
        c_g & s_g \\
        -s_gc_{J_x} & c_g c_{J_x}& c_g s_{J_x} & s_g s_{J_x} \\
        s_g s_{Jx} & -c_g s_{J_x} & c_g c_{J_x} & s_g c_{J_x}\\
         &  & -s_gc_{J_x} & c_g c_{J_x}& c_g s_{J_x} & s_g s_{J_x} \\
         &  & s_g s_{Jx} & -c_g s_{J_x} & c_g c_{J_x} & s_g c_{J_x}\\
         &&&&&&\ddots
    \end{pmatrix}.
\end{align}
Above, we use the shorthand notation: $c_g = \cos(gT), s_g = \sin(gT), c_{J_x} = \cos(J_xT)$ and $s_{J_x} = \sin(J_x T)$. The zero and $\pi$ modes satisfy the eigenvalue equation
\begin{align}
    &K\psi_0 = \psi_0;   &K\psi_\pi = -\psi_\pi,
    \label{Eq: 0PiMode Eigen Eq}
\end{align}
which guarantees commutation and anti-commutation relations with the Floquet unitary. Here, we simply write down the answer which can be checked by direct substitution into \eqref{Eq: 0PiMode Eigen Eq}),
\begin{align}
    \psi_0 \propto  \sum_{l=1}\left\{\left[ \cos\left(\frac{gT}{2}\right)a_{2l-1} + \sin\left(\frac{gT}{2}\right)a_{2l}\right] \right. \nonumber\\
    \left. \times\left[ \tan\left( \frac{gT}{2} \right)\cot\left( \frac{J_xT}{2} \right) \right]^{l-1} \right\}, \label{Eq: Zero Mode}\\
    \psi_\pi \propto  \sum_{l=1}\left\{\left[ \sin\left(\frac{gT}{2}\right)a_{2l-1} - \cos\left(\frac{gT}{2}\right)a_{2l}\right] \right. \nonumber\\ 
    \left. \times\left[ -\cot\left( \frac{gT}{2} \right)\cot\left( \frac{J_xT}{2} \right) \right]^{l-1}\right\}. \label{Eq: Pi Mode}
\end{align}
When applying FGR \eqref{Eq: Fermi's Golden Rule 0Mode} and \eqref{Eq: Fermi's Golden Rule PiMode}, we numerically construct the normalized zero and $\pi$ modes according to the analytic expressions \eqref{Eq: Zero Mode} and \eqref{Eq: Pi Mode}. Note that the commutation and anti-commutation relations only hold in the thermodynamic limit. However, the analytic solutions are localized on the edge with commutation and anti-commutation relations spoiled by a number which is exponentially small in the system size. Hence, one can still apply FGR with a finite system size truncation of \eqref{Eq: Zero Mode} and \eqref{Eq: Pi Mode}.

\section{FGR for the decay of the  infinite temperature autocorrelation} \label{sec:C}

In this appendix, we provide the full derivation of the FGR decay-rate of the infinite temperature autocorrelation of almost strong zero and $\pi$ modes. The full Floquet unitary consists of two parts: one arising from a perturbing interaction $V$ and the other arising from the unperturbed Floquet hamiltonian $H_F^0$,
\begin{align}
    U &= e^{-i V T} e^{-i H_F^0 T}.
\end{align}
To second order in $V$, one obtains
\begin{align}
    U \approx \left(1 - i V T - \frac{V^2 T^2}{2} \right)e^{-i H_F^0 T}.
\end{align}
Now, we consider the time evolution of the zero and $\pi$ modes after one period, $U^\dagger \psi_\eta U$ with $\eta = 0$ or $\pi$ for zero and $\pi$ modes respectively. Up to second order,
\begin{align}
    \psi_\eta(1) = e^{i \mathcal{L}_0T}\left(1 + iT\mathcal{L}_V + T^2\mathcal{G}_{V^2} - \frac{T^2}{2}\mathcal{F}_{V^2}\right)\psi_\eta.
\end{align}
The notations are as follows: $\mathcal{L}_0 \psi_\eta = [H_F^0, \psi_\eta],\ \mathcal{L}_V \psi_\eta = [V,\psi_\eta],\ \mathcal{G}_{V^2} \psi_\eta = V\psi_\eta V$ and $\mathcal{F}_{V^2} \psi_\eta = \{V^2,\psi_\eta\}$. After $N$ periods, 
\begin{align}        
    \psi_\eta(N)=\left[e^{i\mathcal{L}_0T}\left(1 + iT\mathcal{L}_V + T^2\mathcal{G}_{V^2} - \frac{T^2}{2}\mathcal{F}_{V^2}\right)  \right]^N \psi_\eta.
\end{align}
The infinite temperature autocorrelation is given by
\begin{align}
    A_\infty^\eta(N) = \frac{1}{2^L} \text{Tr}[\psi_\eta(N)\psi_\eta].
\end{align}
We will only expand up to second order in $V$ and denote $A_{\infty,n}^{\eta}$ to be the autocorrelation function to $n$-th order in $V$. At the zero-th order, one does not pick up any terms containing $V$, so that 
\begin{align}
    A_{\infty,0}^\eta(N) = \frac{1}{2^L} \text{Tr}\left[\left\{ e^{i\mathcal{L}_0TN} \psi_\eta \right\}\psi_\eta \right] = e^{i\eta N},
\end{align}
where we have applied the commutation relation $e^{i\mathcal{L}_0T} \psi_\eta = e^{i\eta}\psi_\eta$ and employed the normalization $\text{Tr}[\psi_\eta \psi_\eta]/2^L = 1$.

At first order,  $\mathcal{L}_V$ appears once in the expansion
\begin{align}
    &A_{\infty,1}^\eta(N) \nonumber\\ 
    &= \frac{1}{2^L}\sum_{n=0}^{N-1}\text{Tr}\left[ \left\{e^{i\mathcal{L}_0T(N-n)}  (iT\mathcal{L}_V)e^{i\mathcal{L}_0Tn} \psi_\eta\right\} \psi_\eta\right].
\end{align}
With cyclic permutation within the trace, one can show that $\text{Tr}\left[ \left\{e^{i\mathcal{L}_0T}  O_1\right\} O_2\right] = \text{Tr}\left[ O_1 \left\{e^{-i\mathcal{L}_0T}  O_2\right\} \right]$ for arbitrary operators $O_1$ and $O_2$. Also, from the commutation relations,  the first order expansion is further simplified as
\begin{align}
    A_{\infty,1}^\eta(N) = \frac{e^{i\eta N}}{2^L}\sum_{n=0}^{N-1}\text{Tr}\left[ \left\{ (iT\mathcal{L}_V) \psi_\eta\right\} \psi_\eta\right] = 0.
\end{align}
Note that we have used $e^{i\eta} = e^{-i\eta}$ ($e^{2i\eta} = 1$) since we only consider $\eta = 0$ or $\pi$. The above quantity is traceless due to the cyclic property of the trace: $\text{Tr}\left[ \left\{ \mathcal{L}_V \psi_\eta\right\} \psi_\eta\right] = \text{Tr}\left[ \psi_\eta \left\{ -\mathcal{L}_V \psi_\eta\right\} \right] = 0$.

Last, at second order, one has $\mathcal{G}_{V^2}$ once, $\mathcal{F}_{V^2}$ once or $\mathcal{L}_V$ twice in the expansion.
\begin{align}
    &A_{\infty,2}^\eta(N) \nonumber\\
    &= \frac{1}{2^L}\sum_{n=0}^{N-1} \text{Tr} \left[ \left\{e^{i\mathcal{L}_0T(N-n)}  ( T^2 \mathcal{G}_{V^2})e^{i\mathcal{L}_0Tn} \psi_\eta \right\} \psi_\eta \right]\nonumber\\
    &- \frac{1}{2^L}\sum_{n=0}^{N-1} \text{Tr} \left[ \left\{e^{i\mathcal{L}_0T(N-n)}  \left( \frac{T^2}{2} \mathcal{F}_{V^2}\right)e^{i\mathcal{L}_0Tn} \psi_\eta \right\} \psi_\eta \right] \nonumber\\
    &+ \frac{1}{2^L}\sum_{n\geq 1, m\geq0}^{n+m = N-1} \text{Tr}\left[\left\{e^{i\mathcal{L}_0T(N-n-m)}  ( iT\mathcal{L}_V)e^{i\mathcal{L}_0Tn} \right.\right. \nonumber\\
    &\qquad\qquad\qquad\qquad   \left.\left. \times( iT\mathcal{L}_V)e^{i\mathcal{L}_0Tm} \psi_\eta\right\} \psi_\eta\right].
    \label{Eq: 2nd order three parts}
\end{align}
In the first and second trace, one obtains an overall factor $e^{i\eta N}$. Moreover, $\text{Tr}[\{\mathcal{G}_{V^2}\psi_\eta\}\psi_\eta] = \text{Tr}[V\psi_\eta V\psi_\eta]$ and $\text{Tr}[\{\mathcal{F}_{V^2}\psi_\eta\}\psi_\eta] = 2\text{Tr}[\psi_\eta V V\psi_\eta]$. Combining them and summing over $n$ leads to a simple form
\begin{align}
    &\frac{1}{2^L}\sum_{n=0}^{N-1} \text{Tr} \left[ \left\{e^{i\mathcal{L}_0T(N-n)}  ( T^2 \mathcal{G}_{V^2})e^{i\mathcal{L}_0Tn} \psi_\eta \right\} \psi_\eta \right]\nonumber\\
    &- \frac{1}{2^L}\sum_{n=0}^{N-1} \text{Tr} \left[ \left\{e^{i\mathcal{L}_0T(N-n)}  \left( \frac{T^2}{2} \mathcal{F}_{V^2}\right)e^{i\mathcal{L}_0Tn} \psi_\eta \right\} \psi_\eta \right]\nonumber\\
   &= -\frac{e^{i\eta N}NT^2}{2}\times \frac{1}{2^L}\text{Tr}[ \dot{\psi_\eta} \dot{\psi_\eta}],
\end{align}
where we define $\dot{\psi_\eta} = i\mathcal{L}_V \psi_\eta$. Now the last piece is the term with $\mathcal{L}_V$ in \eqref{Eq: 2nd order three parts}. As we have learnt from the first order expansion, $e^{i\mathcal{L}_0 T(N-n-m)}$ and $e^{i\mathcal{L}_0 Tm}$ contribute an overall factor $e^{i\eta (N-n)}$. Then, one associates the first $i\mathcal{L}_V$ with the last $\psi_\eta$ by cyclic permutation. After summing over $m$, one obtains
\begin{align}
    &\frac{1}{2^L}\sum_{n\geq 1, m\geq0}^{n+m = N-1} \text{Tr}\left[\left\{e^{i\mathcal{L}_0T(N-n-m)}  ( iT\mathcal{L}_V)e^{i\mathcal{L}_0Tn} \right.\right. \nonumber\\
    &\qquad\qquad\qquad\qquad \left.\left. \times( iT\mathcal{L}_V)e^{i\mathcal{L}_0Tm} \psi_\eta\right\} \psi_\eta\right]\nonumber\\
    &= -\frac{NT^2}{2^L}\sum_{n\geq 1}^{N}e^{i\eta(N-n)}\left(1-\frac{n}{N}\right) \text{Tr}\left[\dot{\psi_\eta}(n) \dot{\psi_\eta}\right],
\end{align}
where we define $\dot{\psi_\eta}(n) = e^{i\mathcal{L}_0Tn}\dot{\psi_\eta}$.

On combining the above results, the autocorrelation function up to second order in $V$ is
\begin{align}
    A_\infty^\eta (N) \approx A_{\infty,0}^\eta(N) + A_{\infty,2}^\eta(N),
\end{align}
where
\begin{align}
    &A_{\infty,0}^\eta (N) = e^{i\eta N}\\ 
    &A_{\infty,2}^\eta (N)\nonumber\\
    &=  -\frac{e^{i\eta N}NT^2}{2^L} \left( \frac{1}{2}  \text{Tr} \left[ \dot{\psi_\eta} \dot{\psi_\eta} \right] + \sum_{n\geq 1}^{\infty} e^{i\eta n}\text{Tr}\left[\dot{\psi_\eta}(n) \dot{\psi_\eta}\right] \right). \label{Eq: 2nd Order of A}
\end{align}
Note that we approximate the upper bound of the summation $N$ by $\infty$, and therefore the $(1-n/N)$ in the summation is replaced by 1. Since we study quantities where the life-time is long, $N$ is chosen to be a large number. In addition, $\text{Tr}[\dot{\psi_\eta}(n) \dot{\psi_\eta}]$ decays fast with a time scale much smaller than $N$. Therefore, we can simply replace $N$ by $\infty$ in the summation.

The autocorrelation function with decay rate $\Gamma_\eta$ can be formulated as $A_\infty^\eta(N) = e^{i\eta N} e^{-\Gamma_\eta NT} \approx e^{i\eta N}(1-\Gamma_\eta NT)$. By comparing this to the second order expansion, we obtain the FGR decay rate
\begin{align}
    \Gamma_\eta = \frac{T}{2^L} \left( \frac{1}{2}  \text{Tr} \left[ \dot{\psi_\eta} \dot{\psi_\eta} \right] + \sum_{n = 1}^{\infty} e^{i\eta n}\text{Tr}\left[\dot{\psi_\eta}(n) \dot{\psi_\eta}\right] \right),
    \label{Eq: Gamma eta}
\end{align}
which is the first line in \eqref{Eq: Fermi's Golden Rule 0Mode} for $\eta = 0$ and \eqref{Eq: Fermi's Golden Rule PiMode} for $\eta = \pi$. One can reformulate the terms in the round parenthesis as
\begin{align}
    &\frac{1}{2}  \text{Tr} \left[ \dot{\psi_\eta} \dot{\psi_\eta} \right] + \sum_{n\geq 1}^{\infty} e^{i\eta n}\text{Tr}\left[\dot{\psi_\eta}(n) \dot{\psi_\eta}\right] \nonumber\\
    &= \sum_{i,j} |\langle i| \Dot{\psi}_\eta |j\rangle|^2 \left( \frac{1}{2} + \sum_{n=1}^\infty e^{i(\epsilon_i T - \epsilon_j T +\eta)n} \right),
\end{align}
where $|i\rangle$ and $|j\rangle$ are eigenbases of $U_0$. The relation between Dirac delta function and summation of exponential is given by
\begin{align}
    \sum_{m=-\infty}^\infty 2\pi\delta(x+2\pi m) &= \sum_{n=-\infty}^\infty e^{ixn} = 2\left( \frac{1}{2} + \sum_{n=1}^\infty e^{ixn} \right).
\end{align}
Therefore, the FGR decay rate can be expressed as
\begin{align}
    \Gamma_\eta = \frac{1}{2^L} \sum_{i,j} |\langle i| \Dot{\psi}_\eta |j\rangle|^2  \pi \delta_F\left(\epsilon_i -\epsilon_j + \frac{\eta}{T} \right),
\end{align}
where $\delta_F(\epsilon)=\sum_m \delta(\epsilon+m 2 \pi/T)$ is the $\delta$ function encoding energy conservation modulo $2 \pi/T$. The matrix element can be further re-cast as
\begin{align}
    |\langle i| \Dot{\psi}_\eta |j\rangle| = |\langle i| (V - \psi_\eta V \psi_\eta)\psi_\eta |j\rangle| = |\langle i| \Tilde{V}|\Tilde{j}\rangle|,
\end{align}
where we use $\psi_\eta^2 = 1$ and define $|\Tilde{j}\rangle = \psi_\eta|j\rangle$ and $\Tilde{V} = V - \psi_\eta V \psi_\eta$. Finally, we arrive at the results in the second lines of \eqref{Eq: Fermi's Golden Rule 0Mode} and \eqref{Eq: Fermi's Golden Rule PiMode}
\begin{align}
    \Gamma_\eta = \frac{1}{2^L} \sum_{i,j} |\langle i| \Tilde{V} |\Tilde{j}\rangle|^2  \pi \delta_F\left(\epsilon_i -\epsilon_j + \frac{\eta}{T} \right).
\end{align}

\begin{figure*}[t]
    \centering \includegraphics[width=0.32\textwidth]{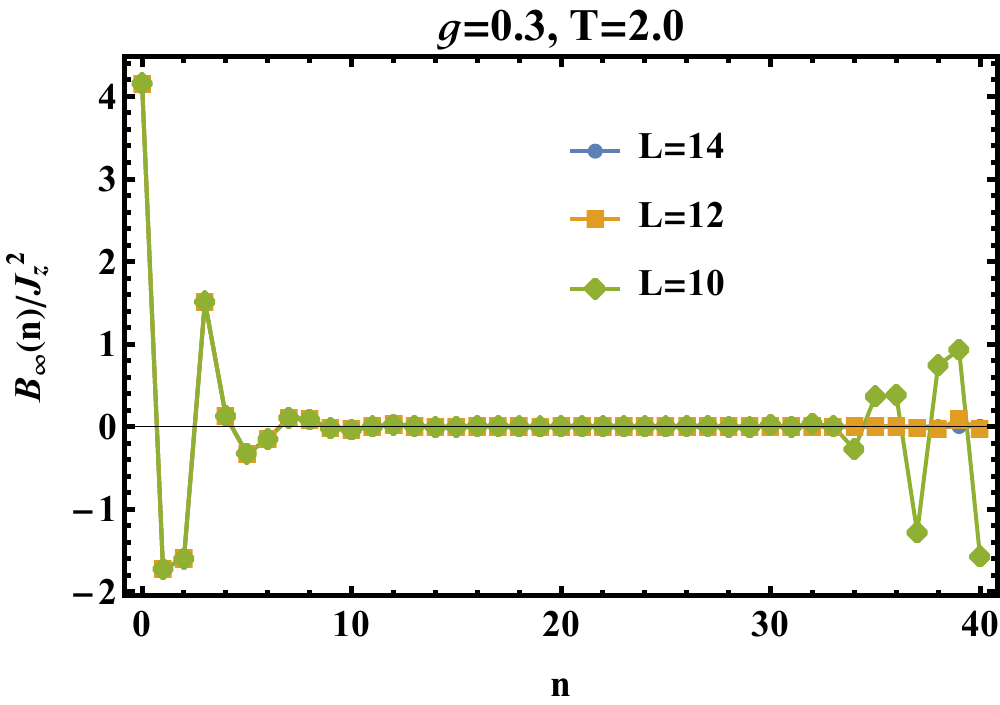} \includegraphics[width=0.32\textwidth]{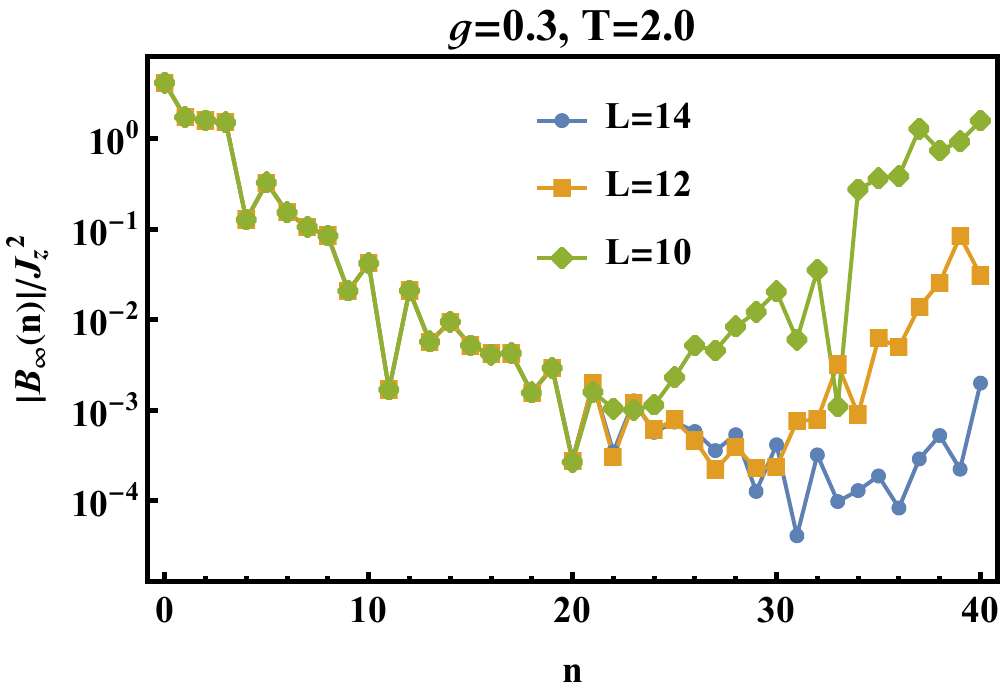} \includegraphics[width=0.32\textwidth]{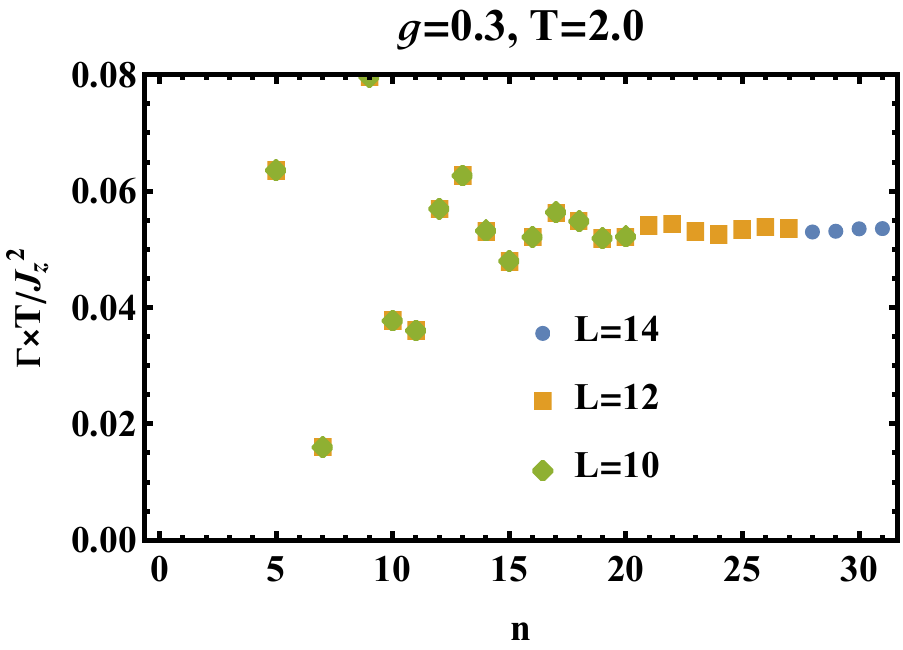}
    \caption{The infinite temperature autocorrelation of $\Dot{\psi_0}$, $B_n = \text{Tr}[\dot{\psi_0}(n) \dot{\psi_0}]/2^L$ (left and middle) and the decay rate (right). $J_z$ is the perturbation strength and the results are $J_z$ independent on multiplying by $1/J_z^2$. The left panel shows a fast decay that supports the approximation in \eqref{Eq: 2nd Order of A}. The fluctuation becomes large at later times for $L=10$, which is the revival effect of a finite size system. This can be clearly seen in the middle panel, where the revivals occur later as the system size increases. To take finite system size into account, we truncate the summation of $n$ in \eqref{Eq: Gamma eta} up to the minimum value in the middle plot.  In the right panel, we show how the decay rate converges as the upper bound of the summation (denoted by $n$ on the $x$-axis) increases. }
    \label{Fig: B autocorrelation}
\end{figure*}

In Fig.~\ref{Fig: B autocorrelation} we use the zero mode as an example to demonstrate the numerical computation of the infinite temperature autocorrelation $B_\infty(n) = \text{Tr}[\dot{\psi_0}(n) \dot{\psi_0}]/2^L$,  and the decay rate derived from it based on \eqref{Eq: Fermi's Golden Rule 0Mode}.

\section{Many particle quasi-energy continuum} \label{sec:D}

In this appendix, we illustrate how to numerically construct the many particle quasi-energy continuum. For fixed parameters $J_x,g,T$, the bulk quasi-energy spectrum $\epsilon_k T$ is a continuum within the interval $[ \epsilon_{{\rm min}}T, \epsilon_{{\rm max}} T]$, where $\epsilon_{{\rm min}} = \text{min}\{ \epsilon_0, \epsilon_\pi \}$ and $\epsilon_{{\rm max}} = \text{max}\{ \epsilon_0, \epsilon_\pi \}$, where $\epsilon_k$ is given in \eqref{Eq: Bulk Dispersion}. In this paper, the perturbation we consider is $J_z\sum_i \sigma_i^z \sigma_{i+1}^z = -J_z \sum_i a_{2i-1}a_{2i}a_{2i+1}a_{2i+2}$, i.e, a four Majorana interaction term. In the language of Feynman diagrams, the decay rate comes from the self-energy  diagram obtained from contracting say $n$ number of four Majorana interaction terms with two external lines left out. At $2n$-th order, there are $4n-1$ internal lines. The resonance condition is numerically determined by constructing the $4n-1$ particle quasi-energy continuum from the bulk dispersion.

Let us start with the 1-particle continuum. Each internal line can represent either the creation or annihilation of a quasi-particle because a  Majorana is a linear combination of the  creation and annihilation operator of a complex  fermion (or Bogoliubov particle).  The 1-particle continuum is therefore: $[ \epsilon_{\text{min}}T, \epsilon_{\text{max}}T ] \cup [ -\epsilon_{\text{max}}T, -\epsilon_{\text{max}}T ]$. For $n = 1$, we have to consider the 3-particle continuum. The possible combinations are: create (annihilate) 3 quasi-particles, create (annihilate) 2 quasi-particles and annihilate (create) 1 quasi-particle. The 3-quasi-particle continuum is therefore $[ 3\epsilon_{\text{min}}T, 3\epsilon_{\text{max}}T ] \cup [ -3\epsilon_{\text{max}}T, -3\epsilon_{\text{max}}T ] \cup[ (2\epsilon_{\text{min}}-\epsilon_{\text{max}})T, (2\epsilon_{\text{max}}- \epsilon_{\text{min}})T]\cup [ (\epsilon_{\text{min}}-2\epsilon_{\text{max}})T, (\epsilon_{\text{max}}-2 \epsilon_{\text{min}})T] $. For larger $n$, the construction is similar, and we do not show it here.

Once all the energy continuum are constructed, one has to fold them into the window $[-\pi,\pi]$ since quasi-energies are only defined modulo  $2\pi$. For a given continuum, $[a,b]$, we shift it into the $[-\pi,\pi]$ interval as follows. First we shift
\begin{align}
    [a,b] \rightarrow [a',b'],
\end{align}
where
\begin{align}
    a' = a - 2\pi\left\lfloor\frac{a+\pi}{2\pi}\right\rfloor,\\
    b' = b - 2\pi \left\lfloor\frac{a+\pi}{2\pi}\right\rfloor,
\end{align}
with the floor function $\lfloor\ \rfloor$. Based on the three possible conditions we further fold $[a',b']$ into the $[-\pi,\pi]$ interval in the following different manners.

If $b' - a' > 2\pi$,
\begin{align}
    [a',b'] \rightarrow [-\pi, \pi].
\end{align}
If $b' > \pi$ and $b' - a' < 2\pi$
\begin{align}
    [a', b'] \rightarrow [a', \pi]\cup[-\pi,b'-2\pi].
\end{align}
If $b' < \pi$ and $b' - a' < 2\pi$,
\begin{align}
    [a', b'] \rightarrow [a', b'].
\end{align}
These three conditions cover all possible cases for a given interval $[a', b']$.

\section{Finite system size effects in numerical results} \label{sec:E}

\begin{figure*}[t]
    \centering      \includegraphics[width=0.32\textwidth]{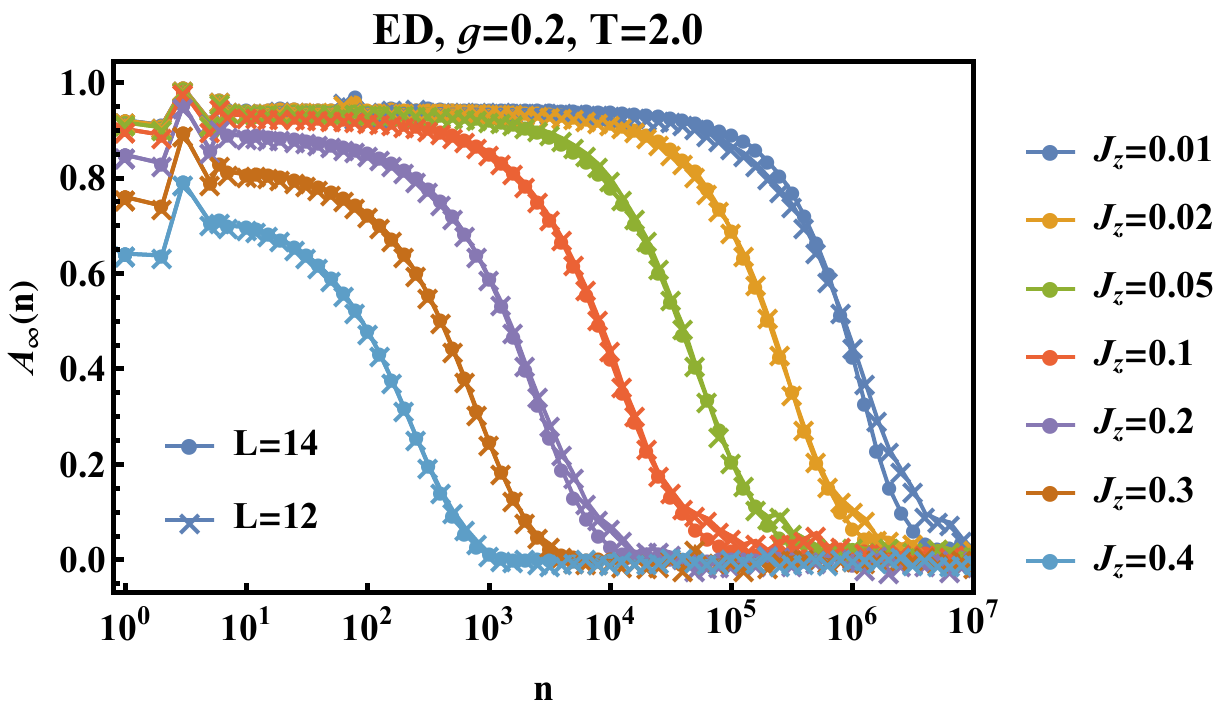} \includegraphics[width=0.32\textwidth]{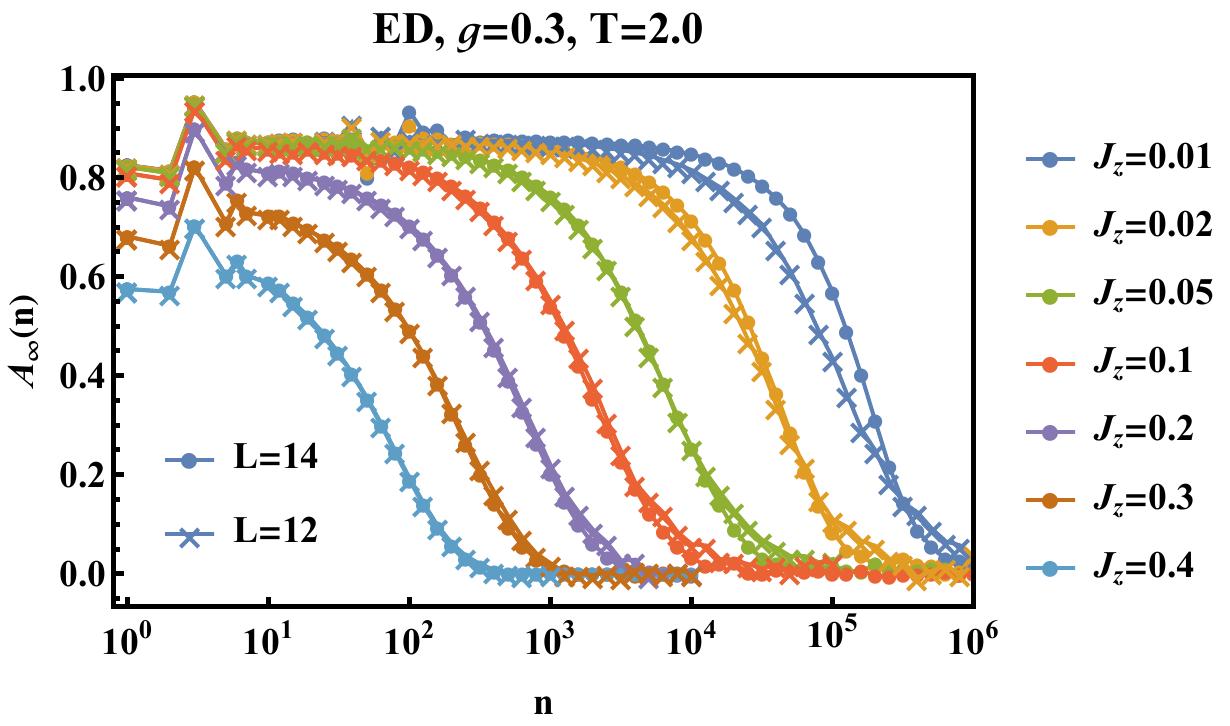} \includegraphics[width=0.32\textwidth]{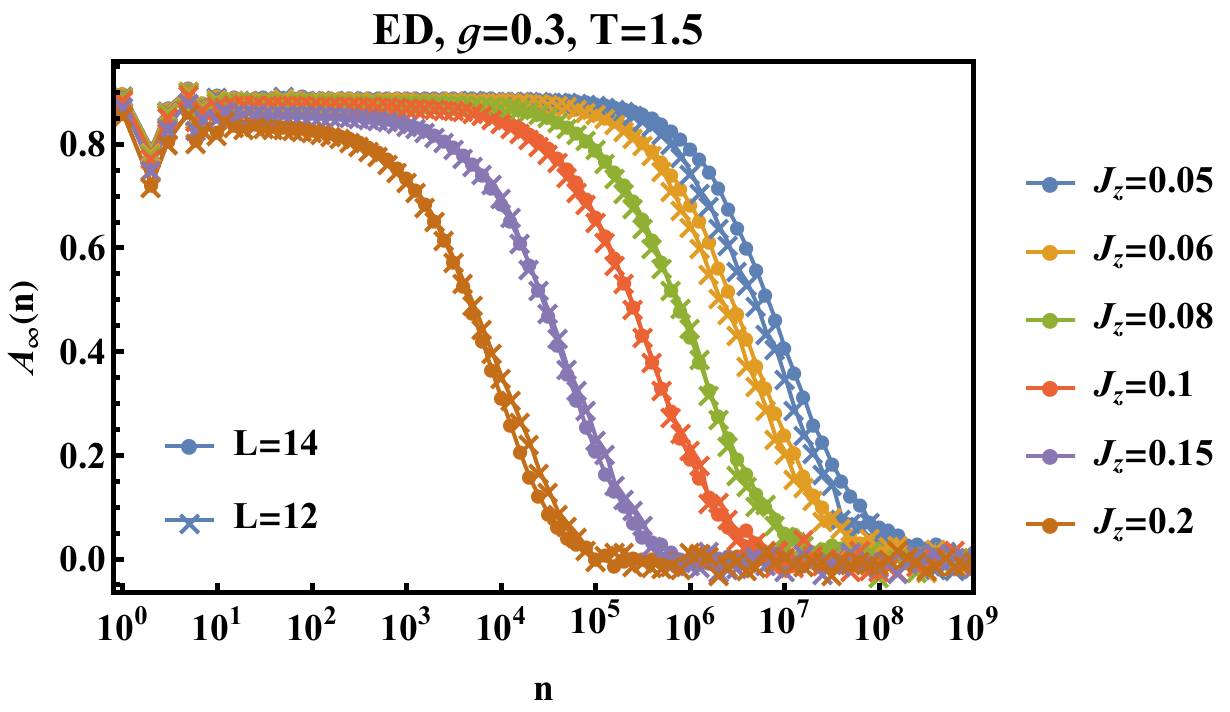} \includegraphics[width=0.32\textwidth]{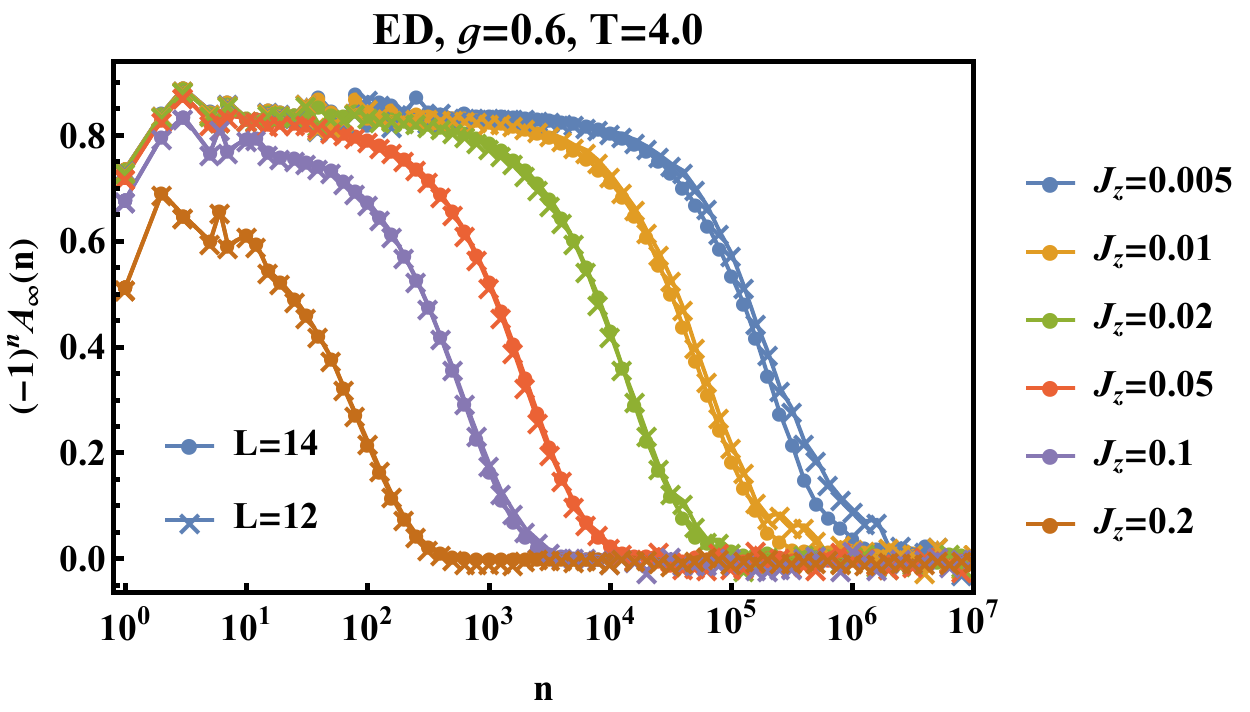}  \includegraphics[width=0.32\textwidth]{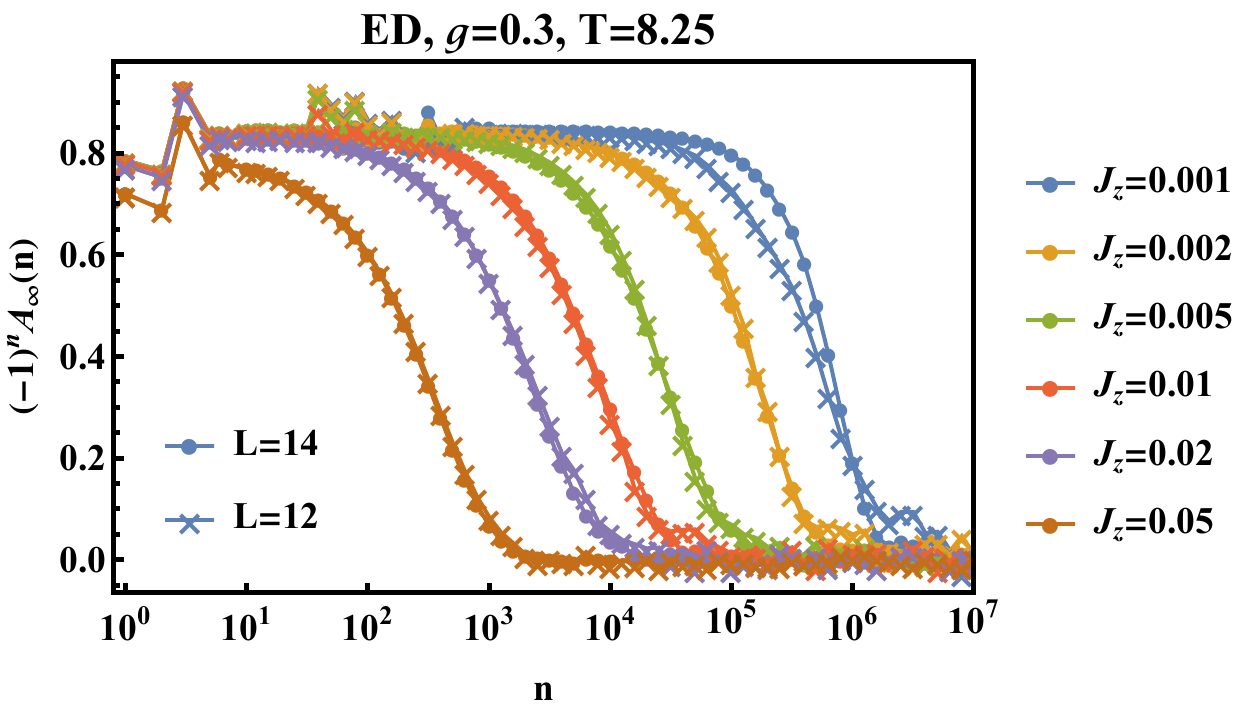} \includegraphics[width=0.32\textwidth]{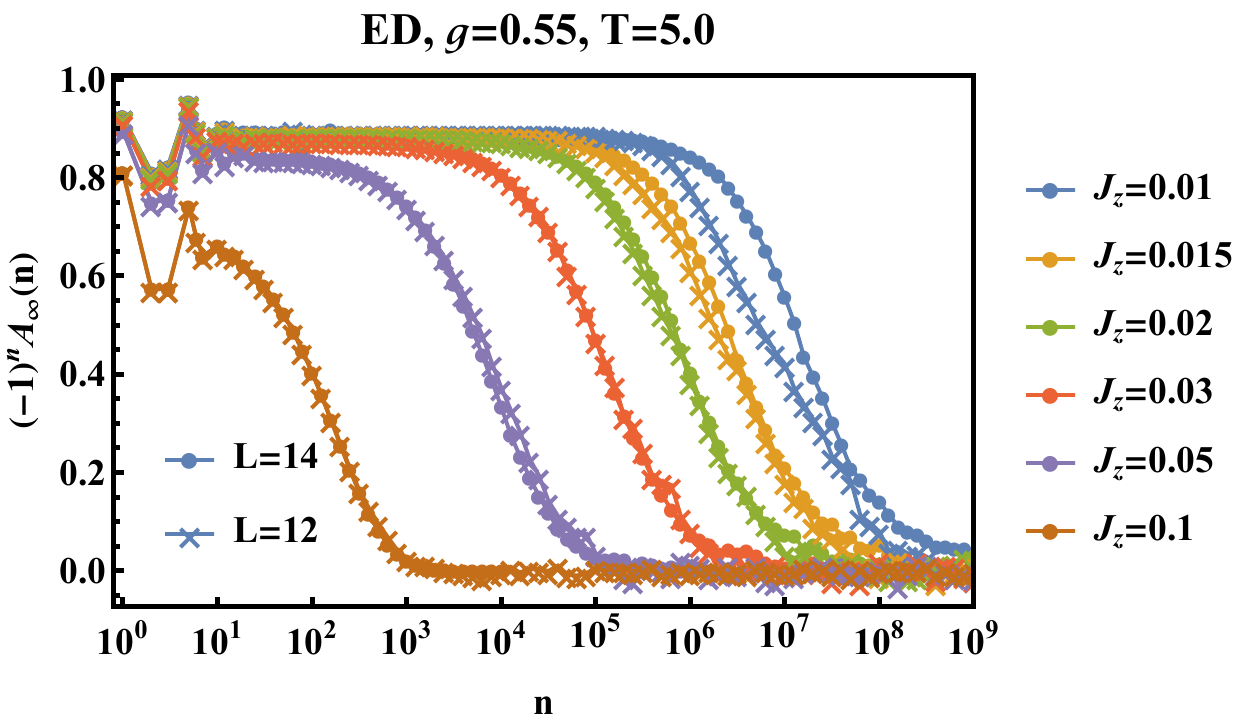}
    \caption{Top panels: Almost strong zero mode. Bottom panels: Almost strong $\pi$ mode. 
    Comparison between the infinite temperature autocorrelation functions for $L=12,14$. The parameters $g,T,J_z$ are chosen to be the same as Figs.~\ref{Fig: 0Mode Autocorrelation} and \ref{Fig: PiMode Autocorrelation}. For second order perturbation (left and middle panels),  $L=12$ shows a slowly decaying tail which is a finite system size effect. $L=14$ also shows finite system size effects which manifest as a deviation from saturation from the thermodynamic limit (non-overlapping of $L=12,14$ plots) for $n \gtrsim 10^6$. For fourth order processes (right panels), the life-times for small $J_z$ are longer than $n=10^6$. At these times, the system size effect is already strong for $L=14$.}
    \label{Fig: L12 vs L14}
\end{figure*}

In numerical computations, we obtain the autocorrelation functions in the thermodynamic limit by increasing the system size until the results saturate. In Fig.~\ref{Fig: L12 vs L14} we show a comparison of the decay of the almost strong modes for two different system sizes, $L=12$ and $L=14$. In the second order region (left and middle panels of Fig.~\ref{Fig: L12 vs L14}), the finite system size effects arise in the tails, in particular, $L=12$ shows a slower decay at late times.

The perfect exponential decay comes from the fact that the energy spectrum is continuous in the thermodynamic limit and hence the delta function condition is obeyed in the FGR formula. However, for finite system sizes, the discrete energy spectrum will be detected at time scales long as compared to the inverse of the energy spacing of the multi-particle excitations. In bare perturbation theory to order $J_z^{2 n}$, the relevant level spacing is proportional to  $1/L^{4 n-1}$. Due to the $4n-1$ bulk energies involved in the scattering process, there are $O(L^{4 n-1})$ multi-particle energies, $\sum_{i=1}^{4n-1} (\pm \epsilon_{k_i})$, entering, e.g., in \eqref{Eq: 0-mode energy conservation} or \eqref{Eq: Pi-mode energy conservation}. Thus, finite size effects are expected when the decay rate $\Gamma\sim W (J_z/W)^{2 n}$ becomes smaller than $W/L^{4 n-1}$ where $W$ is the relevant quasi-particle bandwidth. Thus finite-size effects are expected for $J_z \lesssim W/L^{3/2}$ for $n=1$ (left and middle panels of Fig.~\ref{Fig: L12 vs L14}) and for $J_z \lesssim W/L^{7/4}$ for $n=2$ (right panels of Fig.~\ref{Fig: L12 vs L14}). This is roughly consistent with our numerical results. Note that the estimate above does not take into account that bulk scattering leads to a finite lifetime of bulk modes, which can further suppress finite-size effects 
because the many-particle level spacing is exponentially small in $L$  \cite{mitra2023}.


%

\end{document}